\documentclass[aps,prd,reprint,superscriptaddress,nofootinbib,longbibliography]{revtex4-2}

\usepackage{amsmath,amssymb,bm}
\usepackage{graphicx}
\usepackage[colorlinks=true,
linkcolor=blue,
citecolor=blue,
urlcolor=blue]{hyperref}
\usepackage{booktabs}
\usepackage{xcolor}
\usepackage{mathtools}

\numberwithin{equation}{section}

\newcommand{\be}{\begin{equation}}
\newcommand{\ee}{\end{equation}}
\newcommand{\bea}{\begin{eqnarray}}
\newcommand{\eea}{\end{eqnarray}}
\newcommand{\bse}{\begin{subequations}}
\newcommand{\ese}{\end{subequations}}
\newcommand{\beqa}{\begin{eqnarray}}
\newcommand{\eeqa}{\end{eqnarray}}
\newcommand{\beqar}{\begin{eqnarray*}}
\newcommand{\eeqar}{\end{eqnarray*}}
\newcommand{\bi}{\begin{itemize}}
\newcommand{\ei}{\end{itemize}}
\newcommand{\bn}{\begin{enumerate}}
\newcommand{\en}{\end{enumerate}}

\newcommand{\ba}{\begin{array}}
\newcommand{\ea}{\end{array}}
\newcommand{\bc}{\begin{center}}
\newcommand{\ec}{\end{center}}

\definecolor{darkgreen}{rgb}{0,0.3,0}
\definecolor{darkblue}{rgb}{0,0,0.3}
\definecolor{darkred}{rgb}{0.7,0,0}

\def\lae{\mathrel{\mathop{\smash{\lower .5 ex \hbox{$\stackrel<\sim$}}}}}
\def\lae{\mathrel{\mathop{\smash{\lower .5 ex \hbox{$\stackrel>\sim$}}}}}

\def\doi#1#2{\href{http://doi.org/#1}{#2}}

\begin{document}

\title{Zero modes and oscillatory instabilities of a Lorentz-violating Kalb–Ramond field on a Schwarzschild background}
\author{Hong-Da Lyu}
\email{hongdalyu@sdu.edu.cn}
\affiliation{School of Physics, Southeast University, Nanjing 211189, China}
\affiliation{Key Laboratory of Particle Physics and Particle Irradiation (MOE), Institute of Frontier and Interdisciplinary Science, Shandong University, Qingdao, Shandong 266237, China}
\author{Zhi Xiao}
\email{Corresponding author: spacecraft@pku.edu.cn}
\affiliation{North China Electric Power University, Beijing 102206, China}
\affiliation{Hebei Key Laboratory of Physics and Energy Technology, North China Electric Power University, Baoding 071000, China}
\author{Zhong-Xi Yu}
\email[]{zhongxiyu@yau.edu.cn}
\affiliation{College of Physics and Electronic Information Engineering, Jining Normal University, Wulanchabu, 012000, China}
\author{Shoulong Li}
\email{shoulongli@hunnu.edu.cn}
\affiliation{Department of Physics, Key Laboratory of Low Dimensional Quantum Structures and Quantum Control of Ministry of Education, Institute of Interdisciplinary Studies, Hunan Normal University, Changsha 410081, China}
\affiliation{Hunan Research Center of the Basic Discipline for Quantum Effects and Quantum Technologies, Hunan Normal University, Changsha 410081, China}

\begin{abstract}
We study equilibrium configurations and linear perturbations of a Lorentz-violating Kalb--Ramond field with a quartic symmetry-breaking potential and a nonminimal Riemann coupling on a fixed Schwarzschild background. The electric equilibrium is determined algebraically by a characteristic function that can develop a finite-radius double root. Approaching this configuration, the electric monopole response scales as $|\widehat e(\omega,r_c)|\propto(\gamma_c-\gamma)^{-1/2}$, while the propagating monopole remains regular. For higher multipoles we obtain the exact axial zero-mode tower $\xi_{\ell n}=(\ell+n+1)(\ell+n+2)/3$. At $\ell=1$ a complementary static mode gives a second low-frequency branch. Increasing the positive Riemann coupling causes two purely imaginary growing modes to collide and form an oscillatory unstable pair, with square-root real-part splitting. Independent numerical checks confirm that this behavior persists under changes of the equilibrium parameters. The corrected bare sign removes the monopole ghost, but a constrained high-momentum negative-energy direction remains on the asymptotically electric branch. We therefore distinguish the existence of classical growing modes from the stronger requirement of a positive-energy theory.
\end{abstract}
\maketitle

\section{Introduction}
\label{sec:introduction}
Lorentz symmetry is a fundamental spacetime symmetry underlying both standard model 
of particle physics and general relativity (GR).
However, various candidate theories aiming to unify quantum field theory with GR, 
including bosonic string field theory \cite{SBLV1989a, SBLV1989b}, 
loop quantum gravity \cite{Gambini1998, Alfaro1999}, 
and spacetime foam models \cite{Amelino-Camelia1996, Ellis1999}, 
suggest that Lorentz symmetry might be violated. 
In these scenarios, Lorentz symmetry can be violated either explicitly 
or spontaneously. 
However, spontaneous Lorentz symmetry breaking can elegantly evade the potential inconsistency 
between the Bianchi identity for the Einstein tensor
and the covariant conservation of a generic stress-energy tensor that 
includes contributions from Lorentz violating (LV) background fields \cite{SMEgra}.

In spontaneous Lorentz symmetry breaking (SLSB), 
a dynamical tensor field acquires a nonvanishing vacuum expectation value (VEV), 
thereby selecting preferred directions in the local frame 
while preserving the underlying Lorentz invariance of the action
\cite{Kostelecky:1988zi,Colladay:1996iz,Colladay:1998fq}.
The vector bumblebee models is one of the simplest 
and most extensively studied realizations of this mechanism \cite{BertolamiParamos2005, Bluhm:2007bd,Casana:2017jkc}, 
in which a vector field $B_\mu$ is driven by a suitable potential to a vacuum configuration $\langle B_\mu\rangle=b_\mu$, 
with $b_\mu b^\mu=\mathrm{const.}$. 
In gravitational settings, the bumblebee field may couple non-minimally  to curvature through the $B^\mu B^\nu R_{\mu\nu}$ term, 
allowing the LV vacuum to modify the spacetime geometry.
Consequently, such models have been extensively investigated in various gravitational settings and for their potential observational signatures 
\cite{BaileyKostelecky2006, Bluhm:2007bd, Casana:2017jkc, MalufNeves2021BH, LiangXuLuShao2022, 
Xu2022frb, Luo:2026oxw, Luo:2026duf, JiEtAl2024, Liu:2025oho,Liu:2024axg,Chen:2025ypx,Li:2025tcd}.
Nevertheless, the dynamical consistency of such models remains a subtle issue 
\cite{Bluhm-2008CS, Bonder-2015DA, Seifert-2018C, Seifert-2019SH,Escobar-2022SCC, Bailey-2025BG, ZhuXL2026}.
In particular, the commonly adopted Higgs-like quartic potential does not generically 
lead to a Hamiltonian bounded from below once the full constraint structure is taken into account, 
while related analyses have revealed degeneracies of the constraint algebra 
and potential ambiguities in the evolution near the symmetry-breaking vacuum, and also stability of perturbation around the vacuum
are also discussed
\cite{Bonder-2015DA, Seifert-2018C, Seifert-2019SH, Escobar-2022SCC, Bailey-2025BG, ZhuXL2026, Seturidze:2026daz}.
More recently, the consistency and global regularity of rotating bumblebee vacua have also attracted attention \cite{GuoFan2026}, 
revealing nontrivial obstructions associated with the symmetry and regularity of constant-norm vector configurations.

Beyond vector realizations, SLSB can also arise from higher-form fields,
such as the Kalb–Ramond (KR) field $B_{\mu\nu}=-B_{\nu\mu}$, 
a two-form field that arises naturally in string theory \cite{Kalb:1974yc}. 
In its minimally coupled form and in the absence of a potential, the two-form possesses the gauge symmetry
$B_{\mu\nu}\rightarrow B_{\mu\nu}
+\partial_\mu\Lambda_\nu-\partial_\nu\Lambda_\mu$.
Self-interactions as $\frac{\lambda}{4}(B_{\mu\nu}B^{\mu\nu}-b^2)^2$ can drive the field to a nonzero vacuum
configuration $\langle B_{\mu\nu}\rangle=b_{\mu\nu}$, spontaneously breaking Lorentz invariance. 
The KR field with generic nonminimal gravitational couplings, 
together with the Nambu--Goldstone and massive modes associated with SLSB, was systematically investigated in Ref. \cite{Altschul:2009ae}. 
Further developments include the topological monopole solution formed by the LV KR field itself \cite{Seifert2010Monople}, 
the clarification of the propagating degrees of freedom through spin-projection and propagator analyses \cite{Maluf:2018jwc}, 
the breakdown at one loop of the usual quantum equivalence between dual antisymmetric-tensor and vector descriptions 
in the presence of LV \cite{Buchbinder2008-QE, AashishPanda2019}, 
and the dynamical generation of SLSB through fermion-KR couplings \cite{Assuncao2019DLSB}. 
These studies show that, although the underlying symmetry-breaking mechanism is analogous to that of vector bumblebee models, 
antisymmetric-tensor realizations possess distinctive constraint, topological, propagating, and quantum features.

In the gravitational sector, the general quadratic non-minimal couplings of the KR field to curvature involve the Ricci scalar, 
Ricci tensor, and Riemann tensor \cite{Altschul:2009ae}. 
Related non-minimally coupled two-form models have been explored in inflationary cosmology \cite{Aashish:2018lhv}, 
while the Riemann coupling $B^{\mu\nu}B^{\rho\sigma}R_{\mu\nu\rho\sigma}$
has been investigated in LV Bianchi-I cosmology, 
where it can generate anisotropic expansion \cite{MalufNeves2022KR}. 
This coupling is particularly relevant to Ricci-flat backgrounds, such as Schwarzschild, 
$R=R_{\mu\nu}=0$ whereas $R_{\mu\nu\rho\sigma}\neq0$, 
so it provides a direct interaction between the antisymmetric tensor and the background tidal curvature. 
In the presence of a symmetry-breaking potential, this curvature contribution 
can further shift the local equilibrium of the KR field away from the minimum defined in flat spacetime.

Nonminimal curvature couplings can, however, also affect the dynamical consistency of antisymmetric-tensor theories. 
Hell and Obata recently showed that, for a massive KR field, 
such couplings can render additional modes strongly coupled, 
while Ricci- and Riemann-tensor couplings can produce a runaway instability on homogeneous 
and isotropic backgrounds with a vanishing KR expectation value \cite{HellObata:2026}. 
These results concern a background branch different from the SLSB configuration considered here, 
but they emphasize that the existence of a classical background does not by itself 
guarantee a well-behaved perturbative spectrum.

A separate line of recent work has focused on black-hole geometries generated by a nonzero KR vacuum expectation value, encompassing rotating, electrically and magnetically charged, and regular solutions, as well as extensions to three-dimensional and higher-dimensional spacetimes~\cite{Lessa:2019bgi, Yang2023BH,Duan:2023gng, Yu:2025odj, Lin:2026ewo,Li:2026vqn}. 
Their phenomenology has been explored through geodesic motion, gravitational lensing, black-hole shadows, thermodynamics,
and observational constraints on LV parameters \cite{Atamurotov:2022wsr, Junior:2024ety, Filho:2023ycx, Liu:2026kvm, Ding:2026smj}. 
More recently, quasinormal modes of matter and metric perturbations 
on {slowly rotating} KR black hole backgrounds have also been investigated \cite{Deng:2025atg}. 
Most of these studies therefore probe the spacetime geometry sourced by the KR vacuum 
and the response of other degrees of freedom to that geometry, 
rather than the intrinsic dynamical spectrum of the LV two-form field itself.

This leaves a complementary question: 
how does the Lorentz-violating KR field itself respond dynamically to a curved tidal background? 
To isolate this sector from gravitational backreaction, 
we keep the Schwarzschild geometry fixed and consider a KR two-form with a quartic symmetry-breaking potential 
and a non-minimal Riemann coupling. 
A closely related fixed-background problem was recently studied for a bumblebee vector field \cite{Seturidze:2026daz},
where the static background and spherically symmetric perturbations were controlled by a characteristic function 
and enhanced monopole perturbations appeared near degenerate background configurations. 
The higher-multipole analysis, however, was restricted to static and asymptotic solutions, 
leaving the full finite-frequency spectrum unresolved.


In this work, we extend this analysis to the LV KR two-form on a fixed Schwarzschild background. We determine the equilibrium configurations, exact static modes, and selected low-lying finite-frequency branches. Horizon regularity fixes the magnetic component to a constant, whereas the electric component is algebraic. For negative Riemann coupling the characteristic function can develop a finite-radius double root. The resulting monopole electric response obeys $|\widehat e(\omega,r_c)|\propto(\gamma_c-\gamma)^{-1/2}$ while the propagating magnetic amplitude remains regular.

The higher-multipole sector has a different structure. Its exact axial zero modes occur at $\xi_{\ell n}=(\ell+n+1)(\ell+n+2)/3$. For $\ell=1$, the analytic mode at $\xi=2$ and a complementary static mode at $\xi\simeq2.2931054$ give two distinct dynamical branches for the reference background. With increasing positive coupling, two purely imaginary growing modes collide near $\xi\simeq2.8588608$ and become an oscillatory unstable pair. The real-part splitting follows a square-root law. Thus the finite-frequency spectrum contains information not supplied by the static modes alone.

Positive bare kinetic energy does not, however, establish positivity after all constraints are eliminated. We explicitly distinguish the regular positive-energy monopole from a remaining high-momentum negative-energy direction on the asymptotically electric branch. The results concern classical linear perturbations of the specified quartic model; they do not establish a ghost-free completion.

The remainder of the paper is organized as follows. Section~\ref{sec:setup} specifies the corrected action and field equations. Section~\ref{sec:equilibrium} constructs the static backgrounds and their critical configurations. Section~\ref{sec:perturbations} studies monopoles, static and dynamical higher multipoles, and constrained kinetic energy. Section~\ref{sec:conclusion} summarizes the results. Numerical implementation and convergence checks are collected in Appendix~\ref{app:numerics}.

\section{Setup and perturbative expansion}
\label{sec:setup}

We consider a KR field propagating on a fixed Schwarzschild
background. It is convenient to use Gullstrand--Painlev\'e coordinates
\cite{Martel:2000rn}, in which the metric takes the form
\begin{equation}
\begin{aligned}
    ds^2 &= U(r)dt^2-2\sqrt{\frac{s}{r}}\,dt\,dr
    -dr^2-r^2d\Omega_2^2,\\
    U(r) &= 1-\frac{s}{r}.
\end{aligned}
\label{eq:GP_metric}
\end{equation}
where $s=2GM$ is the Schwarzschild radius. We use the metric signature
$(+,-,-,-)$. The Gullstrand--Painlev\'e coordinates are regular at the
future event horizon $r=s$ and are related to Schwarzschild coordinates
through
\begin{equation}
    dt_{\rm GP}
    =
    dt_{\rm S}
    +
    \frac{\sqrt{s/r}}{U(r)}\,dr.
    \label{eq:GP_transform}
\end{equation}

Throughout this work, the Schwarzschild geometry is treated as a
nondynamical background, and the gravitational backreaction of the
KR field is neglected. This approximation may be understood
as the probe limit of a theory of the form
\begin{equation}
    S_{\rm tot}
    =
    \frac{1}{16\pi G}
    \int d^4x\sqrt{-g}\,R
    +
    \eta S_B,
    \label{eq:probe_action}
\end{equation}
where $\eta$ is an overall normalization of the KR sector.
Because the same factor multiplies the entire two-form action, the
KR equation of motion is independent of $\eta$, whereas its
contribution to the metric equation is proportional to $\eta$.
The limit $\eta\rightarrow0$ therefore keeps the Schwarzschild
geometry fixed without changing the equilibrium configurations or
the linearized KR spectrum studied below.

For finite $\eta$, the fixed-background approximation requires the
stress tensor of the KR configuration to produce only a
small correction to the Schwarzschild geometry. Schematically, in an
orthonormal frame and at radii of order the horizon scale, this
requires
\begin{equation}
    8\pi G\,\eta\,s^2
    \left|T^{(B)}_{\hat a\hat b}\right|
    \ll 1.
    \label{eq:probe_condition}
\end{equation}
We do not evaluate this condition explicitly here, since the metric
is not varied in the present fixed-background analysis. In
particular, the dimensionless parameter choices used below, such as
$\lambda=\bar P_0=1$, specify the KR configuration but
should not be interpreted as fixing the strength of its gravitational
backreaction.

Since the Schwarzschild spacetime is Ricci flat, nonminimal couplings
involving the Ricci tensor or Ricci scalar vanish on the fixed
background, while the Riemann tensor remains nonvanishing. Motivated
by the general nonminimal curvature couplings of Lorentz-violating
antisymmetric-tensor theories \cite{Altschul:2009ae}, we consider the
action
\begin{equation}
\begin{aligned}
    S_B={}&\int d^4x\sqrt{-g}\,\Bigg[
        \frac{1}{12}H_{\lambda\mu\nu}H^{\lambda\mu\nu}
        -\frac{\lambda}{2}
        \left(B_{\mu\nu}B^{\mu\nu}-\gamma\right)^2\\
        &\hspace{2.5cm}
        +\frac{\xi}{4}B^{\mu\nu}B^{\rho\sigma}
        R_{\mu\nu\rho\sigma}\Bigg].
\end{aligned}
\label{KRaction}
\end{equation}
where $B_{\mu\nu}=-B_{\nu\mu}$ is the KR field and
\begin{equation}
    H_{\lambda\mu\nu}
    =
    \nabla_{\lambda}B_{\mu\nu}
    +
    \nabla_{\mu}B_{\nu\lambda}
    +
    \nabla_{\nu}B_{\lambda\mu}
    \equiv
    3\nabla_{[\lambda}B_{\mu\nu]}
    \label{eq:KR_field_strength}
\end{equation}
is its field strength. With the chosen metric signature, $+H^2/12$ gives a positive quadratic velocity term for the spatial two-form components. Positivity after eliminating the electric constraints is checked separately in Sec.~\ref{sec:kineticcheck}. The parameter $\lambda>0$ controls the
symmetry-breaking quartic potential, $\gamma$ specifies its minimum,
and $\xi$ is the nonminimal coupling to the background Riemann tensor.

Our convention for the Riemann tensor is
\begin{equation}
    R^\rho{}_{\sigma\mu\nu}
    =
    \partial_\mu\Gamma^\rho_{\nu\sigma}
    -
    \partial_\nu\Gamma^\rho_{\mu\sigma}
    +
    \Gamma^\rho_{\mu\lambda}
    \Gamma^\lambda_{\nu\sigma}
    -
    \Gamma^\rho_{\nu\lambda}
    \Gamma^\lambda_{\mu\sigma}.
    \label{eq:Riemann_convention}
\end{equation}
This convention is important for fixing the sign of the effects
associated with the coupling $\xi$.

For convenience, we introduce
\begin{equation}
    X
    \equiv
    B_{\mu\nu}B^{\mu\nu}-\gamma.
    \label{eq:X_def}
\end{equation}
Varying Eq.~\eqref{KRaction} with respect to $B_{\mu\nu}$ gives
\begin{equation}
    -\nabla_{\lambda}H^{\lambda\mu\nu}
    -4\lambda X B^{\mu\nu}
    +\xi
    R^{\mu\nu}{}_{\rho\sigma}
    B^{\rho\sigma}
    =
    0.
    \label{KREOM}
\end{equation}
The curvature coupling contributes even though the background is Ricci
flat and will play an essential role in both the equilibrium solution
and its perturbations.

Following the fixed-background perturbative strategy used for the
bumblebee vector field in Ref.~\cite{Seturidze:2026daz}, we expand the
KR field about an equilibrium configuration
$b_{\mu\nu}$,
\begin{equation}
    B_{\mu\nu}
    =
    b_{\mu\nu}
    +
    \epsilon f_{\mu\nu},
    \label{KRexpansion}
\end{equation}
where $\epsilon$ is a small perturbation parameter. Correspondingly,
\begin{equation}
    H_{\lambda\mu\nu}
    =
    h_{\lambda\mu\nu}
    +
    \epsilon q_{\lambda\mu\nu},
\end{equation}
with
\begin{equation}
    h_{\lambda\mu\nu}
    =
    3\nabla_{[\lambda}b_{\mu\nu]},
    \qquad
    q_{\lambda\mu\nu}
    =
    3\nabla_{[\lambda}f_{\mu\nu]}.
    \label{eq:hq_def}
\end{equation}

We define the deviation from the symmetry-breaking vacuum by
\begin{equation}
    \Delta
    \equiv
    b_{\mu\nu}b^{\mu\nu}-\gamma.
    \label{DeltaDef}
\end{equation}
The quantity $X$ then expands as
\begin{equation}
    X
    =
    \Delta
    +
    2\epsilon
    b_{\mu\nu}f^{\mu\nu}
    +
    {\cal O}(\epsilon^2).
    \label{eq:X_expansion}
\end{equation}

At zeroth order in $\epsilon$, Eq.~\eqref{KREOM} gives the equilibrium
equation
\begin{equation}
    -\nabla_{\lambda}h^{\lambda\mu\nu}
    -4\lambda\Delta b^{\mu\nu}
    +\xi
    R^{\mu\nu}{}_{\rho\sigma}
    b^{\rho\sigma}
    =
    0.
    \label{KRbackgroundEOM}
\end{equation}
An important consequence of the Riemann coupling is that the equilibrium field need not satisfy
$b_{\mu\nu}b^{\mu\nu}=\gamma$.
The Schwarzschild curvature instead induces a position-dependent deviation $\Delta$, which will be determined explicitly in the next section.

At first order in $\epsilon$, the perturbation obeys
\begin{equation}
    -\nabla_{\lambda}q^{\lambda\mu\nu}
    -4\lambda\Delta f^{\mu\nu}
    -8\lambda
    \left(
        b_{\rho\sigma}f^{\rho\sigma}
    \right)
    b^{\mu\nu}
    +\xi
    R^{\mu\nu}{}_{\rho\sigma}
    f^{\rho\sigma}
    =
    0.
    \label{KRlinearEOM}
\end{equation}
The linearized dynamics is therefore determined by the background
deviation $\Delta$, the projection $b_{\mu\nu}f^{\mu\nu}$, and the
Riemann coupling. Together with Eq.~\eqref{KRbackgroundEOM}, this equation provides the
starting point for the analysis below.

\section{The Equilibrium Configuration}
\label{sec:equilibrium}

We consider static and spherically symmetric configurations of the
KR field. The general two-form ansatz can be written as
\begin{equation}
    b
    =
    E(r)\,dt\wedge dr
    +
    P(r)\sin\theta\,d\theta\wedge d\phi ,
    \label{eq:background_ansatz}
\end{equation}
where $E(r)$ and $P(r)$ denote the electric- and magnetic-type
components, respectively. In the following we focus on the branch with
$E(r)\neq0$.

For this ansatz, the only independent nonvanishing component of the
field strength $h={\rm d}b$ is
\begin{equation}
    h_{r\theta\phi}
    =
    P'(r)\sin\theta .
    \label{eq:background_h}
\end{equation}
Throughout the paper, a prime denotes differentiation with respect to
$r$, while an overdot denotes differentiation with respect to $t$. In particular, the electric component $E(r)$ does not enter $h$,
since $d[E(r)\,dt\wedge dr]=0$. Its radial profile will therefore be
determined algebraically by the background field equations.

The norm of the equilibrium two-form is
\begin{equation}
    b_{\mu\nu}b^{\mu\nu}
    =
    2\left(
        -E(r)^2
        +
        \frac{P(r)^2}{r^4}
    \right).
    \label{eq:background_norm}
\end{equation}

Substituting the above ansatz into the background equation
\eqref{KRbackgroundEOM}, the $(t,r)$ component gives
\begin{equation}
    \left(
        -4\lambda\Delta
        -
        \frac{2\xi s}{r^3}
    \right)b^{tr}
    =
    0.
\end{equation}
For the branch with $E(r)\neq0$, one therefore obtains
\begin{equation}
    \Delta(r)
    =
    -\frac{\xi s}{2\lambda r^3}.
    \label{eq:Delta_background}
\end{equation}
Thus the equilibrium norm is shifted from its flat-spacetime vacuum
value by the Schwarzschild tidal curvature, with
$\Delta(r)\propto r^{-3}$.

Using Eq.~\eqref{eq:Delta_background}, the $(\theta,\phi)$ component of the background
equation reduces to
\begin{equation}
    \partial_r
    \left(
        \frac{U}{r^2}P'
    \right)
    =
    0.
\end{equation}
and hence
\begin{equation}
    P'
    =
    \frac{Q r^2}{U},
\end{equation}
where $Q$ is an integration constant. Since $U\simeq(r-s)/s$ near
the horizon, a nonzero $Q$ would lead to
$P\sim Qs^3\log|r-s|$. Regularity at the future horizon therefore
requires $Q=0$, so that
\begin{equation}
    P(r)=P_0.
    \label{eq:P_background}
\end{equation}
with $P_0$ constant. The background field strength then vanishes,
$h=db=0$, while the physical magnetic-type component scales as
$P_0/r^2$.

Substituting Eqs.~\eqref{eq:Delta_background} and
\eqref{eq:P_background} into \eqref{eq:background_norm}, we obtain
\begin{equation}
    E(r)^2
    =
    \frac{P_0^2}{r^4}
    +
    \frac{\xi s}{4\lambda r^3}
    -
    \frac{\gamma}{2}.
    \label{eq:E_background}
\end{equation}
In analogy with the characteristic-function description of the
bumblebee equilibrium solution in Ref.~\cite{Seturidze:2026daz}, we
introduce
\begin{equation}
    {\cal C}_{P\xi\gamma}(r)
    =
    \frac{P_0^2}{r^4}
    +
    \frac{\xi s}{4\lambda r^3}
    -
    \frac{\gamma}{2},
    \label{eq:characteristic_function}
\end{equation}
so that
\begin{equation}
    E(r)
    =
    \sigma
    \sqrt{{\cal C}_{P\xi\gamma}(r)},
    \qquad
    \sigma=\pm1.
\end{equation}
The characteristic function ${\cal C}_{P\xi\gamma}(r)$ determines the
allowed equilibrium configurations and also governs the monopole response
near criticality.

Using the Schwarzschild radius $s$ as the length scale, we introduce
the dimensionless variables
\begin{equation}
    \rho
    =
    \frac{r}{s},
    \qquad
    \bar P_0
    =
    \frac{P_0}{s},
    \qquad
    \bar\gamma
    =
    \gamma s^2,
    \qquad
    \bar E
    =
    sE.
    \label{eq:dimensionless_background}
\end{equation}
Eq.~\eqref{eq:E_background} then becomes
\begin{equation}
    \bar E^2
    =
    \frac{\bar P_0^2}{\rho^4}
    +
    \frac{\xi}{4\lambda\rho^3}
    -
    \frac{\bar\gamma}{2}.
    \label{eq:E_dimensionless}
\end{equation}

We now determine when the equilibrium solution is real. Since
$E^2={\cal C}_{P\xi\gamma}$, the characteristic function must satisfy
\begin{equation}
    {\cal C}_{P\xi\gamma}(r)\geq0
\end{equation}
throughout the region of interest. At spatial infinity this immediately
requires $\gamma\leq0$.

For $\xi\geq0$, the characteristic function is monotonic,
\begin{equation}
    {\cal C}'_{P\xi\gamma}(r)
    =
    -\frac{4P_0^2}{r^5}
    -
    \frac{3\xi s}{4\lambda r^4}
    \leq0 ,
\end{equation}
and no finite-radius minimum is present. The situation changes for
$\xi<0$. In this case the $P_0^2/r^4$ term and the curvature-induced
$\xi s/(4\lambda r^3)$ term compete, and
${\cal C}_{P\xi\gamma}$ develops a minimum at
\begin{equation}
    r_c
    =
    -\frac{16\lambda P_0^2}{3\xi s},
    \qquad
    \rho_c
    \equiv
    \frac{r_c}{s}
    =
    -\frac{16\lambda\bar P_0^2}{3\xi}.
    \label{eq:r_c}
\end{equation}
An exterior minimum, $r_c>s$, occurs when
\begin{equation}
    -\frac{16\lambda P_0^2}{3s^2}
    <
    \xi
    <
    0,
    \label{eq:external_rc_condition}
\end{equation}
or equivalently
$-16\lambda\bar P_0^2/3<\xi<0$.

At the minimum,
\begin{equation}
    {\cal C}_{P\xi\gamma}(r_c)
    =
    \frac{\gamma_c-\gamma}{2},
\end{equation}
where
\begin{equation}
    \gamma_c
    =
    -\frac{27\xi^4s^4}
    {32768\lambda^4P_0^6},
    \qquad
    \bar\gamma_c
    \equiv
    \gamma_c s^2
    =
    -\frac{27\xi^4}
    {32768\lambda^4\bar P_0^6}.
    \label{eq:gamma_c}
\end{equation}
When the minimum lies in the region under consideration, reality of the
equilibrium solution therefore requires $\gamma\leq\gamma_c$.

The limiting case $\gamma=\gamma_c$ is special: the minimum of the
characteristic function touches zero,
\begin{equation}
    {\cal C}_{P\xi\gamma_c}(r_c)
    =
    {\cal C}'_{P\xi\gamma_c}(r_c)
    =
    0.
    \label{eq:double_root}
\end{equation}
We refer to this double-root configuration as a critical equilibrium
configuration. Here the term ``critical'' refers only to the degeneracy
of the equilibrium solution and does not imply a thermodynamic phase
transition.

\begin{figure}[tb]
    \centering
    \includegraphics[width=0.95\columnwidth]
    {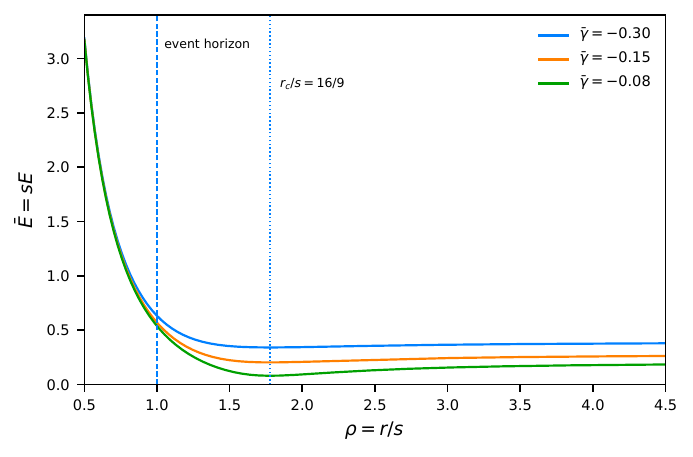}
    \caption{
    Representative equilibrium electric profiles away from the critical
    limit for $\lambda=\bar P_0=1$, $\xi=-3$, and
    $\bar\gamma=-0.30$, $-0.15$, and $-0.08$.
    The positive fixed-sign branch is shown.
    The dashed vertical line marks the future event horizon at
    $\rho=1$, while the dotted line indicates the common minimum at
    $\rho_c=16/9$.
    The continuation into the black-hole interior is shown within the
    fixed-background approximation.
    }
    \label{fig:background_E}
\end{figure}

Figure~\ref{fig:background_E} illustrates several equilibrium profiles
away from the critical limit. We take
$\lambda=\bar P_0=1$ and $\xi=-3$, for which
$\rho_c=16/9$, and vary $\bar\gamma$ while keeping
$\bar\gamma<\bar\gamma_c$. The electric component remains regular
across the future event horizon at $\rho=1$. Its limiting behavior is
also evident from Eq.~\eqref{eq:E_dimensionless}: as $\rho\to0$,
$\bar E\sim|\bar P_0|/\rho^2$, whereas at spatial infinity
$\bar E\to\sigma\sqrt{-\bar\gamma/2}$.

Because $\bar E$ grows toward the Schwarzschild singularity, the
fixed-background description cannot be expected to remain valid
arbitrarily deep inside the black hole. The interior part of
Fig.~\ref{fig:background_E} should therefore be regarded as the formal
continuation of the test-field solution.

\begin{figure}[tb]
    \centering
    \includegraphics[width=0.95\columnwidth]
    {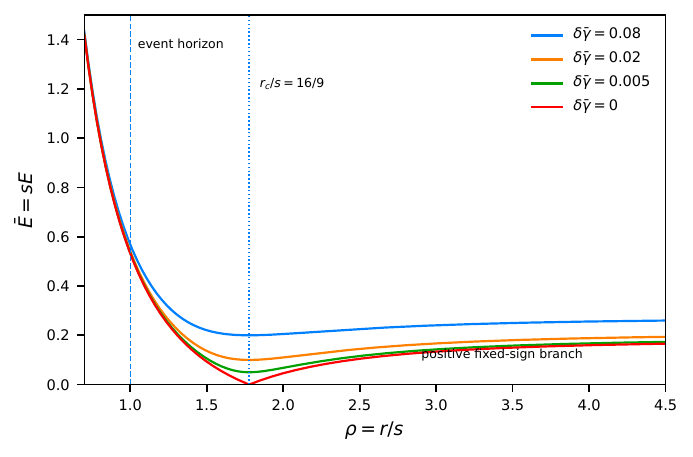}
    \caption{
    Approach to the critical equilibrium configuration for
    $\lambda=\bar P_0=1$ and $\xi=-3$.
    The curves are labeled by
    $\delta\bar\gamma=\bar\gamma_c-\bar\gamma$, with
    $\bar\gamma_c\simeq-0.066742$.
    The dashed vertical line marks the future event horizon at
    $\rho=1$, and the dotted line marks the critical radius
    $\rho_c=16/9$.
    The positive fixed-sign branch, $\sigma=+1$, is shown.
    }
    \label{fig:critical_background_E}
\end{figure}

The approach to the critical configuration is shown in
Fig.~\ref{fig:critical_background_E}. For the same choice
$\lambda=\bar P_0=1$ and $\xi=-3$, one has
\begin{equation}
    \rho_c
    =
    \frac{16}{9},
    \qquad
    \bar\gamma_c
    \simeq
    -0.066742 .
\end{equation}
As $\bar\gamma$ approaches $\bar\gamma_c$ from below, the minimum of
$\bar E(\rho)$ decreases continuously and reaches zero at
$\rho=\rho_c$.

At criticality, the characteristic function has the local expansion
\begin{equation}
    {\cal C}_{P\xi\gamma_c}(r)
    =
    \frac{2P_0^2}{r_c^6}
    (r-r_c)^2
    +
    {\cal O}\left((r-r_c)^3\right).
    \label{eq:C_near_rc}
\end{equation}
For a branch on which the sign of $E$ is kept fixed, this gives
$E(r)\simeq\sqrt{2}|P_0||r-r_c|/r_c^3$, and the electric component
develops a cusp at $r=r_c$. This cusp is not compulsory, however.
Since $E(r)$ enters the background equations algebraically and its
radial derivative does not appear in $h=db$, one may instead continue
through the double root with a change of sign,
$E(r)\simeq\sqrt{2}|P_0|(r-r_c)/r_c^3$. The critical feature is
therefore the double root of ${\cal C}_{P\xi\gamma}$ itself, rather
than the cusp associated with a particular choice of branch.

The finite-radius critical configuration requires both
$P_0\neq0$ and $\xi<0$. It arises from the competition between the
magnetic $r^{-4}$ contribution and the curvature-induced $r^{-3}$
contribution in Eq.~\eqref{eq:characteristic_function}. As we show
next, the same double-root structure also controls the monopole
response near criticality.

\section{Perturbations}
\label{sec:perturbations}

We now analyze solutions of the linearized equation
\eqref{KRlinearEOM} about the equilibrium configuration obtained in
Sec.~\ref{sec:equilibrium}. We take the perturbation amplitude to be
small and require its physical components to vanish at spatial
infinity.

\subsection{Time-Dependent Monopoles}
\label{sec:time_dependent_monopoles}

We first consider time-dependent monopole perturbations. For $\ell=0$,
the most general spherically symmetric perturbation can be written as
\begin{equation}
    f
    =
    e(t,r)\,dt\wedge dr
    +
    p(t,r)\sin\theta\,d\theta\wedge d\phi,
    \label{eq:monopole_ansatz}
\end{equation}
The nonvanishing components of the perturbed field strength
$q={\rm d} f$ are
\begin{equation}
    q_{t\theta\phi}
    =
    \dot p\sin\theta,
    \qquad
    q_{r\theta\phi}
    =
    p'\sin\theta.
    \label{eq:monopole_q}
\end{equation}

Using the equilibrium relation~\eqref{eq:Delta_background}, the
$(t,r)$ component of the linearized equation reduces to the algebraic
constraint
\begin{equation}
    b_{\mu\nu}f^{\mu\nu}
    =
    2\left(
        -E e
        +
        \frac{P_0}{r^4}p
    \right)
    =
    0.
    \label{eq:monopole_constraint}
\end{equation}
The electric perturbation is therefore fixed by
\begin{equation}
    e(t,r)
    =
    \frac{P_0}{r^4E(r)}\,p(t,r).
    \label{eq:e_from_p}
\end{equation}
Thus the monopole dynamics can be formulated entirely in terms of
$p(t,r)$, with $e(t,r)$ reconstructed algebraically from
Eq.~\eqref{eq:e_from_p}.

Substituting the monopole ansatz into the remaining linearized
equations, we obtain
\begin{equation}
    \partial_t
    \left[
        \frac{
            \partial_t p
            -
            \sqrt{s/r}\,\partial_r p
        }{r^2}
    \right]
    -
    \partial_r
    \left[
        \frac{
            \sqrt{s/r}\,\partial_t p
            +
            U\,\partial_r p
        }{r^2}
    \right]
    =
    0.
    \label{eq:gp_monopole_pde}
\end{equation}
Written in Gullstrand--Painlev\'e coordinates, this equation is regular
at the future event horizon. Moreover, the evolution of $p(t,r)$ is
independent of the equilibrium parameters
$\lambda$, $\xi$, $P_0$, and $\gamma$.
Their effect on the monopole perturbation enters through the
reconstruction of $e(t,r)$ in Eq.~\eqref{eq:e_from_p}.

We decompose the perturbation into Fourier modes,
\begin{equation}
    p(t,r)
    =
    \int_{\mathbb R}
    \frac{d\omega}{\sqrt{2\pi}}\,
    e^{-i\omega t}
    \widehat p(\omega,r),
    \label{eq:p_fourier}
\end{equation}
and similarly for $e(t,r)$. For each frequency,
Eq.~\eqref{eq:gp_monopole_pde} becomes
\begin{equation}
    U\widehat p''
    +
    \left(
        U'
        -
        \frac{2U}{r}
        -
        2i\omega\sqrt{\frac{s}{r}}
    \right)\widehat p'
    +
    \left(
        \omega^2
        +
        \frac{5i\omega}{2r}
        \sqrt{\frac{s}{r}}
    \right)\widehat p
    =
    0.
    \label{eq:gp_monopole_ode}
\end{equation}

Regularity at the future horizon fixes the first radial derivative in
terms of the horizon value. Writing
\begin{equation}
    \widehat p(r)
    =
    p_h
    +
    p_1(r-s)
    +
    {\cal O}\left((r-s)^2\right),
\end{equation}
one finds
\begin{equation}
    p_1
    =
    -
    \frac{
        \omega^2+\dfrac{5i\omega}{2s}
    }{
        \dfrac{1}{s}-2i\omega
    }
    p_h .
    \label{eq:monopole_horizon_bc}
\end{equation}
The overall normalization is arbitrary, and we set $p_h=1$ in the
numerical solutions below. We also use the dimensionless frequency
$\bar\omega=s\omega$.

\begin{figure}[tb]
    \centering
    \includegraphics[width=0.95\columnwidth]
    {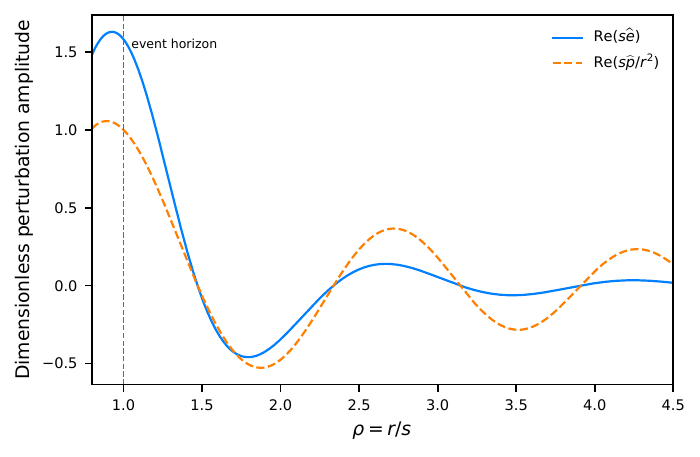}
    \caption{
    Representative horizon-regular monopole perturbation away from
    criticality for
    $\lambda=\bar P_0=1$, $\xi=-3$, $\bar\gamma=-0.30$, and
    $\bar\omega=2\pi$.
    The solid curve shows $\mathrm{Re}(s\widehat e)$ and the dashed curve
    shows the magnetic-type amplitude
    $\mathrm{Re}(s\widehat p/r^2)$.
    The vertical dashed line marks the future event horizon at $\rho=1$.
    The solution is normalized by $\widehat p(s)=1$.
    }
    \label{fig:monopole_noncritical}
\end{figure}

A representative monopole perturbation away from criticality is shown
in Fig.~\ref{fig:monopole_noncritical}. We take
$\lambda=\bar P_0=1$, $\xi=-3$, $\bar\gamma=-0.30$, and
$\bar\omega=2\pi$. Both the propagating amplitude $\widehat p$ and the
electric component reconstructed from Eq.~\eqref{eq:e_from_p} remain
regular across the future event horizon.

We now consider a family of equilibrium backgrounds approaching the
critical configuration by varying $\gamma$ toward $\gamma_c$, while
keeping $s$, $\lambda$, $\xi$, and $P_0$ fixed. In particular, the
location $r_c$ of the minimum remains unchanged along this family.
For the perturbation, we also keep the frequency $\omega$ and the
horizon normalization fixed.

In Fourier space, the algebraic relation~\eqref{eq:e_from_p} becomes
\begin{equation}
    \widehat e(\omega,r)
    =
    \frac{P_0}{r^4E(r)}
    \widehat p(\omega,r).
    \label{eq:e_hat_reconstruction}
\end{equation}
At $r=r_c$, the equilibrium electric component satisfies
\begin{equation}
    E(r_c)^2
    =
    {\cal C}_{P\xi\gamma}(r_c)
    =
    \frac{\gamma_c-\gamma}{2}.
    \label{eq:E_rc_nearcritical}
\end{equation}

The propagation equation~\eqref{eq:gp_monopole_ode} for
$\widehat p$ contains no dependence on $\gamma$. Therefore, for fixed
$\omega$ and fixed horizon normalization,
$\widehat p(\omega,r_c)$ is unchanged as the critical background is
approached. For a generic perturbation with
$\widehat p(\omega,r_c)\neq0$, Eqs.~\eqref{eq:e_hat_reconstruction}
and~\eqref{eq:E_rc_nearcritical} give
\begin{equation}
    \left|
        \widehat e(\omega,r_c)
    \right|
    =
    \frac{\sqrt{2}|P_0|}
    {r_c^4}
    \frac{
        |\widehat p(\omega,r_c)|
    }{
        \sqrt{\gamma_c-\gamma}
    }.
    \label{eq:e_critical_exact}
\end{equation}
Hence the electric response obeys
\begin{equation}
    |\widehat e(\omega,r_c)|
    \propto
    (\gamma_c-\gamma)^{-1/2},
    \qquad
    \gamma\rightarrow\gamma_c^- .
    \label{eq:e_critical_scaling}
\end{equation}
The square-root enhancement therefore applies to generic
fixed-frequency perturbations for which
$\widehat p(\omega,r_c)$ remains finite and nonzero in the critical
limit.

The divergence in Eq.~\eqref{eq:e_critical_scaling} does not arise
from the propagation of $\widehat p$. Rather, it comes from the factor
$1/E(r)$ in the algebraic reconstruction~\eqref{eq:e_hat_reconstruction},
which becomes singular when the minimum of the equilibrium electric
component approaches zero. The monopole enhancement is therefore a
response of the algebraic constraint to the double-root configuration,
rather than a dynamical instability of the propagating mode. A related enhancement of monopole perturbations was found for the
bumblebee vector field in Ref.~\cite{Seturidze:2026daz}. In the present
two-form system, its origin can be traced explicitly to the algebraic
reconstruction factor $1/E(r)$.

\begin{figure}[tb]
    \centering
    \includegraphics[width=0.95\columnwidth]
    {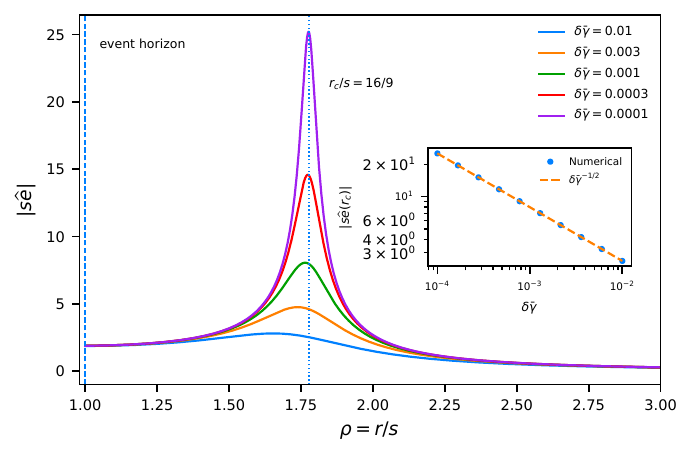}
    \caption{
    Enhancement of the monopole electric response near the critical
    configuration for
    $\lambda=\bar P_0=1$, $\xi=-3$, and $\bar\omega=2\pi$.
    The main panel shows $|s\widehat e|$ for
    $\delta\bar\gamma=\bar\gamma_c-\bar\gamma
    =10^{-2}$, $3\times10^{-3}$, $10^{-3}$,
    $3\times10^{-4}$, and $10^{-4}$.
    The dashed vertical line marks the future event horizon at
    $\rho=1$, and the dotted line marks the critical radius
    $\rho_c=16/9$.
    The inset shows $|s\widehat e(\rho_c)|$ as a function of
    $\delta\bar\gamma$ on logarithmic scales.
    The dashed curve is the analytic scaling
    $|s\widehat e(\rho_c)|\propto
    (\delta\bar\gamma)^{-1/2}$.
    }
    \label{fig:monopole_critical_enhancement}
\end{figure}

This behavior is illustrated in
Fig.~\ref{fig:monopole_critical_enhancement}. We define
$\delta\bar\gamma=\bar\gamma_c-\bar\gamma$ and approach criticality
from $\delta\bar\gamma>0$. The propagating solution $\widehat p$ is
unchanged, whereas the reconstructed electric component develops an
increasingly sharp peak around $\rho=\rho_c$.

At the critical configuration itself,
$|E(r)|\propto|r-r_c|$ near the double root. Therefore, for a generic
perturbation with $\widehat p(\omega,r_c)\neq0$,
Eq.~\eqref{eq:e_hat_reconstruction} gives
\begin{equation}
    |\widehat e(\omega,r)|
    \sim
    \frac{
        |\widehat p(\omega,r_c)|
    }{
        \sqrt{2}\,r_c|r-r_c|
    }.
    \label{eq:e_exact_critical}
\end{equation}
Thus a perturbation that remains finite at $r=r_c$ must satisfy
$\widehat p(\omega,r_c)=0$. A generic horizon-regular perturbation does
not obey this additional condition, so the linear reconstruction of
the electric component breaks down at the exactly critical background.

The static limit provides a simple interpretation of the monopole
perturbation. For $\omega=0$, horizon regularity selects a constant
magnetic perturbation,
$p(r)=p_0$.
Differentiating the equilibrium solution
\eqref{eq:E_background} with respect to $P_0$, while keeping
$s$, $\lambda$, $\xi$, and $\gamma$ fixed, gives
\begin{equation}
    \frac{\partial E(r)}{\partial P_0}
    =
    \frac{P_0}{r^4E(r)}.
\end{equation}
Comparison with the constraint~\eqref{eq:e_from_p} then shows that
\begin{equation}
    f_{\mu\nu}^{(0)}
    =
    p_0
    \frac{\partial b_{\mu\nu}}{\partial P_0}.
    \label{eq:static_tangent_mode}
\end{equation}
Thus the zero-frequency monopole perturbation is simply generated by
an infinitesimal change $P_0\rightarrow P_0+\epsilon p_0$ of the
equilibrium background.

This observation also makes the origin of the critical enhancement
transparent. Since
$\partial E/\partial P_0\propto 1/E$, a small variation of $P_0$
produces a large change in the electric component when
$E(r_c)\rightarrow0$. The static limit therefore provides the same
algebraic picture as the finite-frequency analysis above: the enhanced
electric response originates from the degeneration of the equilibrium
constraint at the double root.

\subsection{Higher-Multipole Perturbations}
\label{sec:higher_multipoles}

We now turn to perturbations with $\ell\geq1$. Unlike the monopole
sector, several two-form polarizations are present and become coupled
once time dependence is included. For convenience, we write the
higher-multipole equations using Schwarzschild time and impose
regularity at the future horizon by transforming to the
Gullstrand--Painlev\'e coordinates introduced in Sec.~\ref{sec:setup}.

We use the standard scalar and vector spherical harmonics on the
two-sphere \cite{Martel:2005ir}. Let $Y_{\ell m}$ satisfy
\begin{equation}
    D^A D_A Y_{\ell m}
    =
    -L Y_{\ell m},
    \qquad
    L\equiv\ell(\ell+1),
\end{equation}
and introduce the associated vector harmonics
\begin{equation}
    Y_A^{\ell m}
    =
    D_A Y_{\ell m},
    \qquad
    X_A^{\ell m}
    =
    \epsilon_A{}^B D_B Y_{\ell m},
\end{equation}
which satisfy
\begin{equation}
    D_A X_B^{\ell m}
    -
    D_B X_A^{\ell m}
    =
    L\epsilon_{AB}Y_{\ell m}.
\end{equation}
For each $(\ell,m)$, the perturbation may be decomposed as
\begin{align}
    f_{tr}
    &=
    a(t,r)Y_{\ell m},
    \nonumber\\
    f_{tA}
    &=
    c(t,r)Y_A^{\ell m}
    +
    h(t,r)X_A^{\ell m},
    \nonumber\\
    f_{rA}
    &=
    d(t,r)Y_A^{\ell m}
    +
    k(t,r)X_A^{\ell m},
    \nonumber\\
    f_{AB}
    &=
    z(t,r)\epsilon_{AB}Y_{\ell m}.
\end{align}
The harmonic indices will be suppressed below.

It is convenient to introduce
\begin{equation}
    F
    =
    a+\dot d-c',
    \qquad
    G
    =
    \dot k-h',
\end{equation}
and
\begin{equation}
    Q
    =
    \dot z-Lh,
    \qquad
    R
    =
    z'-Lk.
\end{equation}
The perturbed field strength is then
\begin{align}
    q_{trA}
    &=
    FY_A+GX_A,
    \nonumber\\
    q_{tAB}
    &=
    Q\epsilon_{AB}Y,
    \nonumber\\
    q_{rAB}
    &=
    R\epsilon_{AB}Y.
\end{align}
The only combination that enters through the symmetry-breaking
potential is
\begin{equation}
    \Sigma
    \equiv
    b_{\mu\nu}f^{\mu\nu}
    =
    2\left(
        -Ea+\frac{P_0}{r^4}z
    \right).
    \label{eq:Sigma_higher}
\end{equation}
We also define
\begin{equation}
    \mu(r)
    =
    \frac{3\xi s}{r^3}.
    \label{eq:mu_higher}
\end{equation}

Using the equilibrium relation~\eqref{eq:Delta_background}, the
independent linearized equations can be written as
\begin{align}
    +\frac{L}{r^2}F
    +8\lambda E\Sigma
    &=
    0,
    \label{eq:higher_1}
    \\
    F'
    -
    \frac{\mu}{U}c
    &=
    0,
    \label{eq:higher_2}
    \\
    \dot F
    -
    \mu U d
    &=
    0,
    \label{eq:higher_3}
    \\
    G'
    -
    \frac{\mu}{U}h
    -
    \frac{Q}{Ur^2}
    &=
    0,
    \label{eq:higher_4}
    \\
    \dot G
    -
    \frac{U}{r^2}R
    -
    \mu U k
    &=
    0,
    \label{eq:higher_5}
    \\
    \frac{\dot Q}{U}
    -
    r^2
    \partial_r
    \left(
        \frac{UR}{r^2}
    \right)
    +
    8\lambda P_0\Sigma
    &=
    0.
    \label{eq:higher_6}
\end{align}
These equations provide the starting point for both the static and
finite-frequency analyses.

\paragraph{Static zero modes.}

We first consider a simple static sector in which only the axial
coefficient $h(r)$ is retained,
\begin{equation}
    a=c=d=k=z=0.
\end{equation}
The full system then reduces consistently to
\begin{equation}
    h''
    -
    \frac{1}{U}
    \left(
        \frac{L}{r^2}
        -
        \frac{3\xi s}{r^3}
    \right)h
    =
    0.
    \label{eq:static_h}
\end{equation}
This equation depends only on $\ell$ and the Riemann coupling $\xi$.

Regularity must be imposed on the two-form itself rather than on a
single Schwarzschild-coordinate component. Since
\begin{equation}
    dt_{\rm S}
    =
    dt_{\rm GP}
    -
    \frac{\sqrt{s/r}}{U}\,dr,
\end{equation}
a horizon-regular static solution must satisfy
$h={\cal O}(U)$ as $r\rightarrow s$.

Introducing
\begin{equation}
    u
    =
    U(r)
    =
    1-\frac{s}{r},
\end{equation}
and writing
\begin{equation}
    h(u)
    =
    u(1-u)^\ell y(u),
\end{equation}
Eq.~\eqref{eq:static_h} becomes
\begin{equation}
    u(1-u)y''
    +
    \left[
        2-(2\ell+4)u
    \right]y'
    -
    \left[
        (\ell+1)(\ell+2)-3\xi
    \right]y
    =
    0.
    \label{eq:h_hypergeometric}
\end{equation}
The solution regular at the future horizon can be written in terms of
the Gauss hypergeometric function,
\begin{equation}
    h(u)
    =
    C_{\ell m}
    u(1-u)^\ell
    {}_2F_1(a,b;2;u),
\end{equation}
with
\begin{equation}
    a
    =
    \ell+\frac32
    +
    \frac12\sqrt{1+12\xi},
    \qquad
    b
    =
    \ell+\frac32
    -
    \frac12\sqrt{1+12\xi}.
\end{equation}

At spatial infinity, a generic solution contains both a decaying and a
growing contribution. The growing part is absent when the
hypergeometric series terminates, which requires
\begin{equation}
    b=-n,
    \qquad
    n=0,1,2,\ldots .
\end{equation}
The allowed couplings are therefore discrete,
\begin{equation}
    \xi_{\ell n}
    =
    \frac{
        (\ell+n+1)(\ell+n+2)
    }{3},
    \qquad
    \ell\geq1,
    \quad
    n\geq0.
    \label{eq:xi_static}
\end{equation}
At these values of $\xi$, the exact zero-frequency solutions are
\begin{equation}
    h_{\ell n}(r)
    =
    C_{\ell m}
    U(r)
    \left(
        \frac{s}{r}
    \right)^\ell
    {}_2F_1
    \left(
        -n,
        2\ell+n+3;
        2;
        U(r)
    \right).
    \label{eq:h_exact}
\end{equation}
The lowest member of the tower is
\begin{equation}
    \xi_{10}
    =
    +2,
    \qquad
    h_{10}(r)
    \propto
    \frac{s(r-s)}{r^2}.
    \label{eq:h10}
\end{equation}

\begin{figure}[tb]
    \centering
    \includegraphics[width=0.95\columnwidth]
    {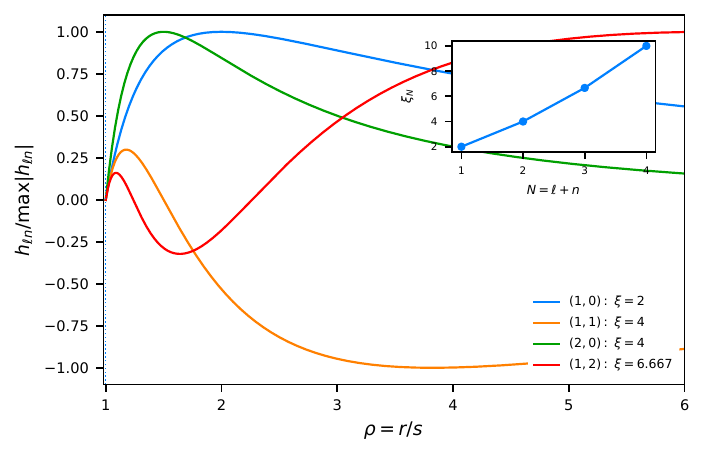}
    \caption{
    Representative exact static higher-multipole zero modes.
    The main panel shows the normalized radial profiles
    $h_{\ell n}/\max|h_{\ell n}|$ as functions of
    $\rho=r/s$ for
    $(\ell,n)=(1,0)$, $(1,1)$, $(2,0)$, and $(1,2)$.
    The corresponding couplings are
    $\xi=2$, $4$, $4$, and $20/3$.
    The vertical dotted line marks the future event horizon at
    $\rho=1$.
    The inset shows the discrete spectrum
    $\xi_N=(N+1)(N+2)/3$, with $N=\ell+n$; the modes
    $(1,1)$ and $(2,0)$ illustrate the degeneracy at $\xi=+4$.
    }
    \label{fig:static_zero_modes}
\end{figure}

The spectrum depends only on $N=\ell+n$,
\begin{equation}
    \xi_N
    =
    \frac{(N+1)(N+2)}{3},
\end{equation}
so different multipoles may occur at the same value of $\xi$. For
example, $(\ell,n)=(1,1)$ and $(2,0)$ are both supported at
$\xi=+4$. Figure~\ref{fig:static_zero_modes} shows several members of
this tower.

These solutions are genuine zero-frequency modes of the complete
linearized system. They do not, however, exhaust the static solution
space. To identify the remaining static modes, we first introduce the
variables used below for the full frequency-dependent problem.

\paragraph{First-order formulation and the second static sector.}

For a mode with time dependence $e^{-i\omega t}$, we define
\begin{equation}
    F
    =
    -\mu\Phi,
    \qquad
    K
    =
    Uk,
\end{equation}
together with
\begin{equation}
    A
    =
    \frac{L}{16\lambda E^2r^2},
    \qquad
    W
    =
    \frac{L}{r^2}-\mu,
\end{equation}
and
\begin{equation}
    D
    =
    L-\mu r^2
    =
    L-\frac{3\xi s}{r}.
\end{equation}
The radial problem then reduces to the six first-order equations
\begin{align}
    \Phi'
    &=
    \frac{3}{r}\Phi
    -
    \frac{c}{U},
    \label{eq:dyn_phi}
    \\
    c'
    &=
    \mu(1-A)\Phi
    +
    \frac{\omega^2}{U}\Phi
    +
    \frac{P_0}{Er^4}z,
    \label{eq:dyn_c}
    \\
    h'
    &=
    -G
    -
    \frac{i\omega}{U}K,
    \label{eq:dyn_h}
    \\
    G'
    &=
    -\frac{W}{U}h
    -
    \frac{i\omega}{Ur^2}z,
    \label{eq:dyn_G}
    \\
    z'
    &=
    -\frac{i\omega r^2}{U}G
    +
    \frac{D}{U}K,
    \label{eq:dyn_z}
    \\
    K'
    &=
    -\frac{i\omega}{U}h
    +
    \frac{3}{r}K
    -
    \frac{P_0L}{Er^4}\Phi.
    \label{eq:dyn_K}
\end{align}
For $\xi\ne0$, these evolution equations contain no inverse powers of $D$. In particular, a
finite-radius zero of $D(r)$ for positive $\xi$ is not a singularity of
the perturbation equations.

At $\omega=0$, the system separates into two independent sectors. The
pair $(h,G)$ reproduces the axial equation~\eqref{eq:static_h}, while
the complementary sector is governed by
\begin{align}
    \Phi'
    &=
    \frac{3}{r}\Phi
    -
    \frac{c}{U},
    \label{eq:static_second_phi}
    \\
    c'
    &=
    \mu(1-A)\Phi
    +
    \frac{P_0}{Er^4}z,
    \label{eq:static_second_c}
    \\
    z'
    &=
    \frac{D}{U}K,
    \label{eq:static_second_z}
    \\
    K'
    &=
    \frac{3}{r}K
    -
    \frac{P_0L}{Er^4}\Phi.
    \label{eq:static_second_K}
\end{align}
The remaining perturbation coefficients are fixed algebraically; in
particular,
\begin{equation}
    d=0,
    \qquad
    a
    =
    \frac{P_0}{Er^4}z
    -
    \mu A\Phi .
    \label{eq:static_second_a}
\end{equation}
Thus this mode is not contained in the purely axial ansatz used to
derive Eq.~\eqref{eq:static_h}.

Regularity of the full two-form at the future horizon requires
$c={\cal O}(U)$ and $K={\cal O}(U)$, while $\Phi$ and $z$ may remain
finite. Together with the condition that the physical perturbation
vanish at spatial infinity, Eqs.~\eqref{eq:static_second_phi}--
\eqref{eq:static_second_K} define a separate static boundary-value
problem. For $\ell=1$, $\lambda=\bar P_0=1$, and
$\bar\gamma=-0.30$, the zero mode continuously connected to the tracked branch occurs at
\begin{equation}
    \xi_0
    \simeq
    2.2931054.
    \label{eq:xi0}
\end{equation}
As we show below, this solution is the zero-frequency endpoint of a
second dynamical branch, which we denote by branch B. The analytic
mode at $\xi_{10}=2$ and the numerical mode at $\xi_0$ therefore have
different static origins.

The character of this second zero mode can also be seen directly from
its radial profile. Normalizing the horizon amplitude to $\Phi_h=1$,
the static matching calculation gives
\begin{equation}
    z_h\simeq0.043474.
    \label{eq:branchB_zh}
\end{equation}
The reconstructed perturbation has nonvanishing $a$, $c$, $k$, and
$z$ components, while $h=G=d=0$. Branch B is therefore not contained
in the purely axial truncation used to obtain
Eq.~\eqref{eq:static_h}; it is a mixed two-form mode of the
complementary static sector. Numerical details and a representative
radial profile are given in Appendix~\ref{app:numerics}.

\paragraph{Finite-frequency spectrum.}

We now impose the black-hole boundary conditions for generic nonzero
$\omega$. At the future horizon, an ingoing mode with the convention
$e^{-i\omega t}$ behaves in Schwarzschild coordinates as
$e^{-i\omega t}U^{-i\omega s}$
\cite{Berti:2009kk}. We therefore factor $U^{-i\omega s}$ from each
radial function and require the remaining amplitudes to be regular at
$U=0$.

Substituting the corresponding near-horizon expansions into
Eqs.~\eqref{eq:dyn_phi}--\eqref{eq:dyn_K}, the leading terms give
\begin{equation}
    c_h=i\omega\Phi_h,
    \qquad
    K_h=h_h,
\end{equation}
together with
\begin{equation}
    (L-3\xi)h_h
    +
    i\omega
    \left(
        z_h-s^2G_h
    \right)
    =
    0.
    \label{eq:horizon_dynamic}
\end{equation}
Thus three of the six leading horizon amplitudes are fixed in terms of
the other three. For generic nonzero $\omega$, the space of ingoing
solutions at the future horizon is therefore three dimensional.

At spatial infinity we select the three outgoing asymptotic solutions.
For ${\rm Im}\,\omega>0$, these modes decay exponentially. We integrate
a basis of the ingoing subspace outward from the horizon and a basis of
the outgoing subspace inward from large radius. Matching the two
three-dimensional subspaces at an intermediate radius $r_m$ gives the
Evans-type eigenvalue condition
\cite{Sandstede2002}
\begin{equation}
    {\cal E}(\omega,\xi)
    =
    \det
    \left[
        {\cal Y}_{\rm H}(r_m),
        {\cal Y}_{\infty}(r_m)
    \right]
    =
    0.
    \label{eq:Evans}
\end{equation}
The asymptotic basis and numerical convergence tests are described in
Appendix~\ref{app:numerics}. The static crossings were checked with a separate collocation problem; finite-frequency roots were tested under endpoint and matching-radius variations. The original-sign collocation data are not reused.

We now specialize to the lowest multipole, $\ell=1$, and use $s$ as
the unit of length, with
\begin{equation}
    \lambda
    =
    \bar P_0
    =
    1,
    \qquad
    \bar\gamma
    =
    -0.30.
    \label{eq:dynamic_parameters}
\end{equation}
With the convention $e^{-i\omega t}$, a mode with
${\rm Im}\,\omega>0$ grows exponentially.

The low-frequency spectrum contains two distinct branches. Branch A passes through the exact axial zero mode at $\xi_{10}=2$ and grows for $\xi>2$. Its eigenfunction approaches $h_{10}$ at the crossing. Branch B emerges from the complementary static solution at $\xi_0\simeq2.2931054$. The ordering of these endpoints differs from that obtained with a negative bare kinetic term. At the branch-B static crossing, branch A is already unstable:
\begin{equation}
 \omega_A s\simeq0.13215113\,i
 \quad\text{at}\quad\xi=\xi_0.
 \label{eq:GammaB_minus2}
\end{equation}
Thus an exact zero-frequency solution does not imply that every other mode is stable at that coupling.

Continuing both branches toward larger positive coupling, the two growing frequencies collide at
\begin{equation}
 \xi_{\rm m}\simeq2.85886084,\qquad
 \omega_{\rm m}s\simeq0.16261649\,i.
 \label{eq:mode_merger}
\end{equation}
For $\xi>\xi_{\rm m}$ they leave the imaginary axis. Reality of the time-domain equations preserves the symmetry $\omega\mapsto-\omega^*$, giving
\begin{equation}
 \omega_\pm=\pm\omega_R+i\omega_I,\qquad\omega_I>0.
 \label{eq:complex_pair}
\end{equation}
For example, at $\xi=3$ we obtain $\omega_\pm s\simeq\pm0.03073337+0.17493256i$. Near the merger,
\begin{equation}
 |\operatorname{Re}(\omega_\pm s)|\simeq A_{\rm m}\sqrt{\xi-\xi_{\rm m}},
 \qquad A_{\rm m}\simeq0.08853.
 \label{eq:sqrt_splitting}
\end{equation}
This local square-root behavior is consistent with a branch-point collision~\cite{Heiss:2012dx}. We do not infer eigenvector coalescence from a double frequency root alone, and therefore retain the term mode collision rather than asserting an exceptional point.

\begin{figure}[tb]
\centering
\includegraphics[width=0.95\columnwidth]{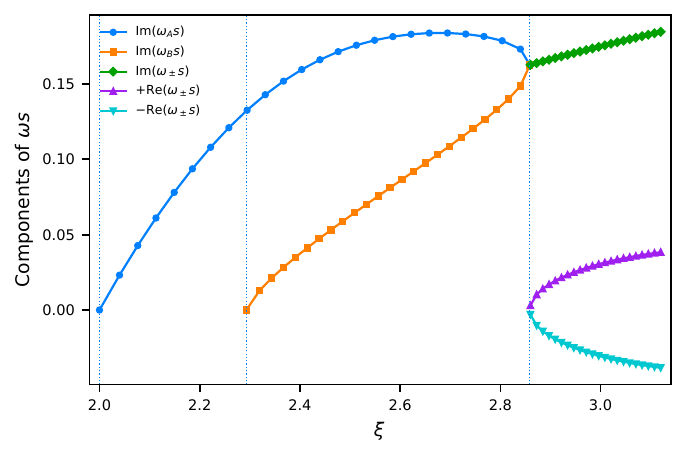}
\caption{Recomputed low-lying $\ell=1$ spectrum for $\lambda=\bar P_0=1$ and $\bar\gamma=-0.30$ with positive bare kinetic energy. Circles and squares show branches A and B, diamonds the common imaginary part of the off-axis pair, and triangles the two real parts. Dotted lines mark $\xi_{10}=2$, $\xi_0\simeq2.2931054$, and $\xi_{\rm m}\simeq2.8588608$. Markers are computed frequencies or independently established static endpoints; lines guide the eye. The plot does not include the complete damped spectrum.}
\label{fig:dynamical_instability}
\end{figure}

We repeat the computation for $\bar\gamma=-0.20$ and for $\bar P_0=0.95,1.05$. The distinct static crossings, collision of growing modes, and off-axis continuation survive these changes. The quantitative dependence is appreciable and must be recomputed rather than inferred from a sign reflection. Tables~\ref{tab:robustness_gamma} and \ref{tab:robustness_P0} summarize the results. These checks follow the specified branches and do not prove spectral completeness or stability in unsearched regions.

\begin{table}[tb]\centering
\caption{Positive-sign $\ell=1$ spectrum at $\lambda=\bar P_0=1$. The square-root amplitude is a local extrapolation.}
\label{tab:robustness_gamma}
\begin{tabular}{ccccc}\hline\hline
$\bar\gamma$ & $\xi_0$ & $\xi_{\rm m}$ & $\operatorname{Im}(\omega_{\rm m}s)$ & $A_{\rm m}$\\\hline
$-0.30$ & $2.293105$ & $2.858861$ & $0.1626165$ & $0.08853$\\
$-0.20$ & $2.395060$ & $3.139294$ & $0.1858792$ & $0.07975$\\
\hline\hline\end{tabular}\end{table}
\begin{table}[tb]\centering
\caption{Dependence on the magnetic background at $\lambda=1$, $\bar\gamma=-0.30$. Each row was recalculated with the corrected equations.}
\label{tab:robustness_P0}
\begin{tabular}{cccc}\hline\hline
$\bar P_0$ & $\xi_0$ & $\xi_{\rm m}$ & $\operatorname{Im}(\omega_{\rm m}s)$\\\hline
$0.95$ & $2.281505$ & $2.787949$ & $0.1522377$\\
$1.00$ & $2.293105$ & $2.858861$ & $0.1626165$\\
$1.05$ & $2.304761$ & $2.929898$ & $0.1725671$\\
\hline\hline\end{tabular}\end{table}
The limit $\bar P_0\rightarrow0$ is, however, special. At exactly
$P_0=0$, the complementary static system
Eqs.~\eqref{eq:static_second_phi}--\eqref{eq:static_second_K}
separates into independent $(\Phi,c)$ and $(z,K)$ sectors. In
particular,
\begin{equation}
    \Phi=c=K=0,
    \qquad
    z={\rm const},
\end{equation}
is a regular static solution for arbitrary $\xi$. Its physical angular
component decays as $r^{-2}$ at spatial infinity. Thus the
$P_0\rightarrow0$ limit contains a continuous static zero-mode
degeneracy and is not a regular limit of the isolated branch-B
crossing found at nonzero $P_0$. A nonzero magnetic background couples
this mode to the remaining amplitudes and reduces the static
zero-frequency condition to discrete values of $\xi$.

The coexistence of static and growing modes also occurs at higher multipoles. At $\ell=2$ the axial solution is $h_{20}\propto U(s/r)^2$ at $\xi_{20}=4$. A separate positive-sign complementary static mode is found at $\xi\simeq2.1062742$, following the static solution continuously in $L$ from $L=2$ to $L=6$. This numerical continuation identifies a solution, not a proof that it is the lowest root of the complete static spectrum. A different initial guess can converge to a higher static root, so the branch identification matters.

At $\xi=4$, where the axial mode is static, a separately computed growing mode has $\omega s\simeq0.20504875i$. At $\xi=4.02$, we obtain two growing frequencies, $0.00852362i$ and $0.20677809i$. Thus the coexistence result survives the sign correction, although the old numerical values and the identification of individual higher branches cannot simply be carried over. We leave the complete higher-multipole connectivity for future work.

\paragraph{Relation to strong coupling.}

It is useful to distinguish the instabilities found above from the
strong-coupling behavior recently discussed for nonminimally coupled
massive KR fields \cite{HellObata:2026}. In that analysis,
Ricci- and Riemann-tensor couplings were shown to affect the
dynamical degree-of-freedom structure, and runaway modes were found
for a spectator KR field on homogeneous and isotropic
backgrounds with a vanishing background field.

The present problem differs in several respects. Here the
KR field is expanded about a nonzero, static
Lorentz-breaking configuration supported by a quartic potential, and
the spacetime metric is treated as nondynamical. The metric sector of
the fully gravity-coupled theory is therefore absent from the present
perturbation problem. The unstable modes found here are instead global
solutions of the KR boundary-value problem, regular at the
future horizon and satisfying the outgoing condition at spatial
infinity.

These observations do not by themselves establish that all of the
KR modes considered here remain weakly coupled. Conversely,
the existence of a growing eigenmode does not by itself demonstrate
strong coupling. A direct comparison would require an analysis of
the perturbative action about the present nonzero background,
including the kinetic matrix, canonical normalization, and the scale
at which nonlinear interactions become important. We therefore
interpret the results above as classical linear instabilities of the
fixed-background problem, while leaving their possible relation to
strong coupling for future investigation.

\subsection{Constrained kinetic energy}
\label{sec:kineticcheck}
For the monopole, eliminating $e=P_0p/(Er^4)$ gives, up to a positive angular normalization,
\begin{equation}
 S_0^{(2)}=\int dt\,dr\left[\frac{\dot p^2}{2Ur^2}-\frac{U(p')^2}{2r^2}\right].
\end{equation}
The exterior velocity Hessian is positive. This explains why the sign correction removes the bare monopole ghost while leaving its classical propagation equation unchanged.

Other constraints require a separate check. In the asymptotic region the magnetic background vanishes as $P_0/r^2$, while the electric background approaches $E_\infty=\sqrt{-\gamma/2}$. Introduce local electric and magnetic variables $a_i=B_{0i}$ and $b_i=\epsilon_{ijk}B_{jk}/2$. The positive kinetic term is $\tfrac12(\dot{\boldsymbol b}-\boldsymbol\nabla\times\boldsymbol a)^2-\tfrac12(\boldsymbol\nabla\cdot\boldsymbol b)^2$, up to the curl convention. Take the electric condensate along $z$ and a Fourier wave vector along $x$. A magnetic fluctuation along $y$ couples to the nondynamical electric fluctuation along $z$. With real mode amplitudes $q$ and $a$, the quadratic Lagrangian is
\begin{align}
 \mathcal L_k^{(2)}&=\tfrac12(\dot q+ka)^2-\tfrac12m_E^2a^2,\\
 m_E^2&=16\lambda E_\infty^2=-8\lambda\gamma>0.
\end{align}
Eliminating $a$ away from $k^2=m_E^2$ yields
\begin{equation}
 \mathcal L_{k,\mathrm{red}}^{(2)}
 =\frac12\frac{m_E^2}{m_E^2-k^2}\dot q^2.
\end{equation}
The reduced coefficient is negative for $k>m_E$; at equality the constraint elimination is singular. Localized short-wavelength packets far from the black hole retain this negative direction because curvature and magnetic-background corrections vanish there. This is distinct from the low-multipole spectral calculation and does not identify a particular plotted collision mode as the same high-momentum fluctuation. It does show that positive bare kinetic energy alone is insufficient for a healthy quartic theory. Hamiltonian pathologies in Lorentz-breaking tensor models likewise require attention to constraints~\cite{Seifert-2019SH}.

\section{Conclusion}
\label{sec:conclusion}
We have studied static equilibria and linear perturbations of a Lorentz-violating KR field with positive bare kinetic energy on a fixed Schwarzschild background. The magnetic equilibrium remains constant, and the electric component is determined by its characteristic function. For negative Riemann coupling, an exterior minimum can become a double root. The electric monopole response then grows as $(\gamma_c-\gamma)^{-1/2}$, while the propagating magnetic amplitude remains regular. The enhancement is an algebraic constraint effect and should not itself be identified with a dynamical instability or a strong-coupling scale.

With the corrected kinetic sign, the exact axial zero modes occur at
\begin{equation}
 \xi_{\ell n}=\frac{(\ell+n+1)(\ell+n+2)}3.
\end{equation}
Their profiles and degeneracies retain the hypergeometric structure, but their couplings are positive. The complementary static problem is not obtained by reflecting its old eigenvalues. At the reference parameters its tracked dipole root is $\xi_0\simeq2.2931054$, distinct from the analytic axial root at $\xi=2$. The finite-frequency continuation shows that branch A is already growing when branch B reaches its static crossing.

The two purely imaginary growing modes collide at $\xi_{\rm m}\simeq2.85886084$ and $\omega_{\rm m}s\simeq0.16261649i$. For larger coupling, they form the pair $\omega_\pm=\pm\omega_R+i\omega_I$ with $\omega_I>0$. Their real parts open as $0.08853\sqrt{\xi-\xi_{\rm m}}$ locally. The behavior survives changes of both the asymptotic electric parameter and magnetic background. At $\ell=2$, a growing mode also coexists with the exact axial static solution at $\xi=4$. These results establish that the oscillatory instability survives the correction of the bare kinetic sign; its location and the ordering of the static endpoints do change.

The equilibrium critical configuration and the axial zero modes now occupy different coupling signs. This separation should be kept explicit: the critical electric response at negative coupling and the collision shown at positive coupling are two effects of the same model, not successive stages of one parameter sweep through the displayed reference configurations.

The energy interpretation remains qualified. The corrected monopole has positive kinetic energy, but eliminating a nondynamical electric component on the asymptotically electric branch produces a negative kinetic coefficient at sufficiently large spatial momentum. The present quartic theory is therefore not rendered ghost free by the bare sign correction. Its classical equilibria and growing solutions can be computed consistently, while a healthy completion would require further changes to the potential or constraints and a new analysis of the spectrum.

Including gravitational backreaction and metric perturbations remains an important extension. It would determine how these modes mix with gravitational degrees of freedom, but it cannot be assumed to cure the asymptotic matter-sector negative-energy direction. A complete classification of the higher-multipole spectrum, a direct eigenvector test at collisions, and a controlled analysis of nonlinear interactions are also needed before drawing conclusions about a fully coupled and stable theory.

\subsection*{Acknowledgments}
This work is supported by National Natural Science Foundation of China Grants No. 12105098, No. 12447134 and No. 12481540179. 
Z. Xiao express his gratitude to the discussion with Tianbo Liu and the hospitality of Institute of Frontier 
and Interdisciplinary Science of Shandong University.
H-D.L.~is also supported by Postdoctoral Innovation Project of Shandong Province SDCX-ZG-202503036. 
S.L. is supported in part by the National Natural Science Foundation of China (No. 12105098, No.
12481540179) and the Natural Science Foundation of Hunan Province (No. 2022JJ40264),
and the innovative research group of Hunan Province under Grant No. 2024JJ1006, and by
the Excellent Young Scholars Program of the Hunan Provincial Department of Education
under Grant No. 25B0092. Z.X.Y is also supported by the Young Scholars Startup Fund of Jining Normal University.

\vspace{0.95cm}
\appendix

\section{Numerical implementation and convergence tests}
\label{app:numerics}

In this appendix we provide further details of the numerical
calculations discussed in Sec.~\ref{sec:higher_multipoles}. We first
describe the ingoing and outgoing solution spaces used for the
finite-frequency matching problem. We then discuss the complementary
static branch-B zero mode and present representative convergence tests
for the low-lying $\ell=1$ spectrum.

For a fixed frequency $\omega$, the radial equations
\eqref{eq:dyn_phi}--\eqref{eq:dyn_K} form a first-order system for
\begin{equation}
    {\bf Y}
    =
    (\Phi,c,h,G,z,K)^T.
    \label{eq:appendix_Y}
\end{equation}

\subsection{Near-horizon ingoing solutions}
\label{app:horizon_basis}

With the Fourier convention $e^{-i\omega t}$, an ingoing mode behaves
near the future horizon as
\begin{equation}
    e^{-i\omega t}U^{-i\omega s}.
\end{equation}
We therefore factor the universal radial behavior
$U^{-i\omega s}$ and expand the remaining amplitudes regularly around
$U=0$.

As discussed in Sec.~\ref{sec:higher_multipoles}, the leading horizon
coefficients satisfy
\begin{equation}
    c_h=i\omega\Phi_h,
    \qquad
    K_h=h_h,
\end{equation}
and
\begin{equation}
    (L-3\xi)h_h
    +
    i\omega
    \left(
        z_h-s^2G_h
    \right)
    =
    0.
    \label{eq:appendix_horizon_constraint}
\end{equation}
For generic nonzero $\omega$, three horizon amplitudes are therefore
independent. A convenient choice is
$(\Phi_h,h_h,G_h)$, in terms of which
\begin{equation}
    c_h=i\omega\Phi_h,
    \qquad
    K_h=h_h,
    \qquad
    z_h
    =
    s^2G_h
    +
    \frac{i(L-3\xi)}{\omega}h_h .
    \label{eq:appendix_horizon_data}
\end{equation}

Three independent ingoing solutions can thus be generated by choosing
successively
\begin{equation}
    (\Phi_h,h_h,G_h)
    =
    (1,0,0),
    \qquad
    (0,1,0),
    \qquad
    (0,0,1).
\end{equation}
The corresponding Frobenius expansions are evaluated at
\begin{equation}
    r_{\rm h}
    =
    s(1+\epsilon),
    \qquad
    \epsilon\ll1,
\end{equation}
and integrated outward. The residual dependence on $\epsilon$ is
examined below.

\subsection{Outgoing solutions at spatial infinity}
\label{app:infinity_basis}

We next derive the asymptotic basis used at spatial infinity. The
finite-frequency calculations in the main text are performed for
$\gamma<0$, for which
\begin{equation}
    E(r)
    =
    E_\infty
    +
    {\cal O}(r^{-3}),
    \qquad
    E_\infty
    =
    \sigma
    \sqrt{-\frac{\gamma}{2}}.
    \label{eq:E_infinity}
\end{equation}
The remaining background quantities behave as
\begin{align}
    \frac{1}{U}
    &=
    1+\frac{s}{r}
    +
    {\cal O}(r^{-2}),
    \nonumber\\
    \mu
    &=
    {\cal O}(r^{-3}),
    \nonumber\\
    A
    &=
    {\cal O}(r^{-2}),
    \nonumber\\
    W
    &=
    \frac{L}{r^2}
    +
    {\cal O}(r^{-3}),
    \nonumber\\
    D
    &=
    L
    +
    {\cal O}(r^{-1}).
    \label{eq:asymptotic_coefficients}
\end{align}
The terms proportional to $P_0/E$ couple the two sectors only at
subleading order and therefore do not modify the leading outgoing
exponents.

It is useful to define
\begin{equation}
    Z
    \equiv
    \frac{z}{r^2}.
    \label{eq:Z_def}
\end{equation}
The $(\Phi,c)$ sector then takes the asymptotic form
\begin{equation}
    \frac{d}{dr}
    \begin{pmatrix}
        \Phi\\
        c
    \end{pmatrix}
    =
    \left[
    \begin{pmatrix}
        0 & -1\\
        \omega^2 & 0
    \end{pmatrix}
    +
    \frac{1}{r}
    \begin{pmatrix}
        3 & -s\\
        \omega^2s & 0
    \end{pmatrix}
    +
    {\cal O}(r^{-2})
    \right]
    \begin{pmatrix}
        \Phi\\
        c
    \end{pmatrix}.
    \label{eq:asymptotic_polar_system}
\end{equation}
We use the ansatz
\begin{equation}
    \begin{pmatrix}
        \Phi\\
        c
    \end{pmatrix}
    =
    e^{kr}r^\beta
    \left[
        {\bf v}_0
        +
        \frac{{\bf v}_1}{r}
        +
        {\cal O}(r^{-2})
    \right].
    \label{eq:polar_asymptotic_ansatz}
\end{equation}
At leading order,
\begin{equation}
    k^2+\omega^2=0.
\end{equation}
For an outgoing solution we choose
\begin{equation}
    k=+i\omega.
\end{equation}
The next order gives
\begin{equation}
    \beta
    =
    \frac{3}{2}
    +
    i\omega s.
    \label{eq:beta_polar}
\end{equation}
Thus the $(\Phi,c)$ sector supplies one independent outgoing solution.

For the remaining variables, the asymptotic equations may be written
as
\begin{equation}
    \frac{d}{dr}
    \begin{pmatrix}
        h\\
        G\\
        Z\\
        K
    \end{pmatrix}
    =
    \left[
        {\cal M}_0
        +
        \frac{{\cal M}_1}{r}
        +
        {\cal O}(r^{-2})
    \right]
    \begin{pmatrix}
        h\\
        G\\
        Z\\
        K
    \end{pmatrix},
    \label{eq:asymptotic_axial_system}
\end{equation}
where
\begin{equation}
    {\cal M}_0
    =
    \begin{pmatrix}
        0 & -1 & 0 & -i\omega\\
        0 & 0 & -i\omega & 0\\
        0 & -i\omega & 0 & 0\\
        -i\omega & 0 & 0 & 0
    \end{pmatrix},
\end{equation}
and
\begin{equation}
    {\cal M}_1
    =
    \begin{pmatrix}
        0 & 0 & 0 & -i\omega s\\
        0 & 0 & -i\omega s & 0\\
        0 & -i\omega s & -2 & 0\\
        -i\omega s & 0 & 0 & 3
    \end{pmatrix}.
\end{equation}
The leading characteristic equation is
\begin{equation}
    \det
    \left(
        k{\bf 1}-{\cal M}_0
    \right)
    =
    (k^2+\omega^2)^2.
    \label{eq:axial_characteristic}
\end{equation}
The outgoing root $k=+i\omega$ is therefore a double root. Its
eigenspace is one dimensional, so the leading matrix is defective at
this root. The algebraic powers are determined by carrying the
asymptotic expansion to the next order. The corresponding solvability
condition is
\begin{equation}
    \left(
        \beta-i\omega s
    \right)
    \left(
        \beta-\frac32-i\omega s
    \right)
    =
    0.
    \label{eq:axial_beta_condition}
\end{equation}
Hence this sector supplies two independent outgoing channels,
\begin{equation}
    \beta_2
    =
    i\omega s,
    \qquad
    \beta_3
    =
    \frac32+i\omega s.
    \label{eq:axial_betas}
\end{equation}

It is useful to display the leading asymptotic solutions explicitly.
We define the common outgoing-wave factor
\begin{equation}
    {\cal W}_{\rm out}(r)
    =
    e^{i\omega r}
    \left(
        \frac{r}{s}
    \right)^{i\omega s}.
    \label{eq:Wout}
\end{equation}
Since the Schwarzschild tortoise coordinate behaves as
\begin{equation}
    r_*
    =
    r
    +
    s
    \ln
    \left(
        \frac{r}{s}-1
    \right),
\end{equation}
one has
\begin{equation}
    {\cal W}_{\rm out}(r)
    \sim
    e^{i\omega r_*}
\end{equation}
at large $r$, up to an irrelevant constant normalization.

The outgoing mode associated with the $(\Phi,c)$ sector may be chosen
as
\begin{equation}
\begin{gathered}
    \begin{pmatrix}
        \Phi\\
        c
    \end{pmatrix}_{\!\rm out}^{(1)}
    \\[-3pt]
    =
    \left(
        \frac{r}{s}
    \right)^{3/2}
    {\cal W}_{\rm out}
    \left[
        \begin{pmatrix}
            1\\
            -i\omega
        \end{pmatrix}
        +
        \frac{1}{r}
        \begin{pmatrix}
            -\dfrac{3i}{2\omega}\\
            0
        \end{pmatrix}
        +
        {\cal O}(r^{-2})
    \right].
    \end{gathered}
    \label{eq:outgoing_channel_1}
\end{equation}

The first outgoing mode in the $(h,G,Z,K)$ sector is
\begin{equation}
    \begin{pmatrix}
        h\\
        G\\
        Z\\
        K
    \end{pmatrix}_{\!\rm out}^{(2)}
    =
    {\cal W}_{\rm out}
    \left[
        \begin{pmatrix}
            -1\\
            0\\
            0\\
            1
        \end{pmatrix}
        +
        \frac{1}{r}
        \begin{pmatrix}
            -\dfrac{3i}{\omega}\\
            -3\\
            3\\
            0
        \end{pmatrix}
        +
        {\cal O}(r^{-2})
    \right],
    \label{eq:outgoing_channel_2}
\end{equation}
while the second is
\begin{equation}
\begin{gathered}
    \begin{pmatrix}
        h\\
        G\\
        Z\\
        K
    \end{pmatrix}_{\!\rm out}^{(3)}
    \\[-3pt]
    =
    \left(
        \frac{r}{s}
    \right)^{3/2}
    {\cal W}_{\rm out}
    \left[
        \begin{pmatrix}
            -1\\
            0\\
            0\\
            1
        \end{pmatrix}
        +
        \frac{1}{r}
        \begin{pmatrix}
            -\dfrac{3i}{2\omega}\\
            0\\
            0\\
            0
        \end{pmatrix}
        +
        {\cal O}(r^{-2})
    \right].
    \end{gathered}
    \label{eq:outgoing_channel_3}
\end{equation}
The components not displayed in
Eqs.~\eqref{eq:outgoing_channel_1}--\eqref{eq:outgoing_channel_3}
are induced only at subleading orders by the coupling between the two
sectors.

Although the two modes
\eqref{eq:outgoing_channel_2} and
\eqref{eq:outgoing_channel_3} share the same leading eigenvector of
${\cal M}_0$, they are linearly independent because they carry
different algebraic powers of $r$.

Combining the two sectors, the asymptotic solution space therefore
contains three independent outgoing modes for generic nonzero
$\omega$. For the unstable modes studied in the main text,
${\rm Im}\,\omega>0$, and
\begin{equation}
    |{\cal W}_{\rm out}|
    \propto
    e^{-{\rm Im}(\omega)r},
\end{equation}
so all three outgoing channels decay exponentially at spatial
infinity.

The above expansion assumes $\gamma<0$ and generic
$\omega\neq0$, as in the finite-frequency calculations of the main
text. The exactly static problem is treated separately in
Sec.~\ref{sec:higher_multipoles}.

\subsection{Matching condition}
\label{app:matching}

We integrate the three ingoing basis solutions outward from
$r_{\rm h}$ and the three outgoing basis solutions inward from a finite
outer boundary $r_{\max}$. At an intermediate matching radius $r_m$,
we assemble the corresponding solution matrices,
\begin{equation}
    {\cal Y}_{\rm H}(r_m)
    =
    \left(
        {\bf Y}_{\rm H}^{(1)},
        {\bf Y}_{\rm H}^{(2)},
        {\bf Y}_{\rm H}^{(3)}
    \right),
\end{equation}
and
\begin{equation}
    {\cal Y}_{\infty}(r_m)
    =
    \left(
        {\bf Y}_{\infty}^{(1)},
        {\bf Y}_{\infty}^{(2)},
        {\bf Y}_{\infty}^{(3)}
    \right).
\end{equation}
A nontrivial solution satisfying both boundary conditions exists when
the two three-dimensional subspaces have a nonvanishing intersection.
This is equivalent to
\begin{equation}
    {\cal E}(\omega,\xi)
    =
    \det
    \left[
        {\cal Y}_{\rm H}(r_m),
        {\cal Y}_{\infty}(r_m)
    \right]
    =
    0,
    \label{eq:appendix_Evans}
\end{equation}
which is the Evans-type matching condition used in the numerical
calculation \cite{Sandstede2002}.

The zeros of ${\cal E}$ are followed continuously as $\xi$ is varied. In the numerical implementation the state is scaled to $(\Phi,c,h,rG,z/r,K)$, and each propagated three-dimensional basis is periodically orthonormalized. This avoids loss of linear independence when growing and decaying channels have different scales. A nonsingular three-row minor fixes the basis chart locally; its normalization changes the determinant but not its zeros. We also monitor the smallest singular value of the matching matrix with orthonormalized columns.

For the growing modes, the horizon subspace is initialized from the exact residue matrix and the outer subspace from the three decaying eigenvectors of the finite-radius coefficient matrix. The latter prescription approaches the outgoing subspace under inward propagation. The asymptotic expansion above identifies its physical boundary condition. Finite-cutoff errors are assessed by moving both endpoints; small matching residuals alone are not used as estimates of these errors. The static calculation uses a separate boundary-value collocation method. We do not claim an independent finite-frequency collocation verification of the revised spectrum.

\subsection{Static branch-B zero mode}
\label{app:static_branchB}
\begin{figure}[tb]\centering
\includegraphics[width=0.95\columnwidth]{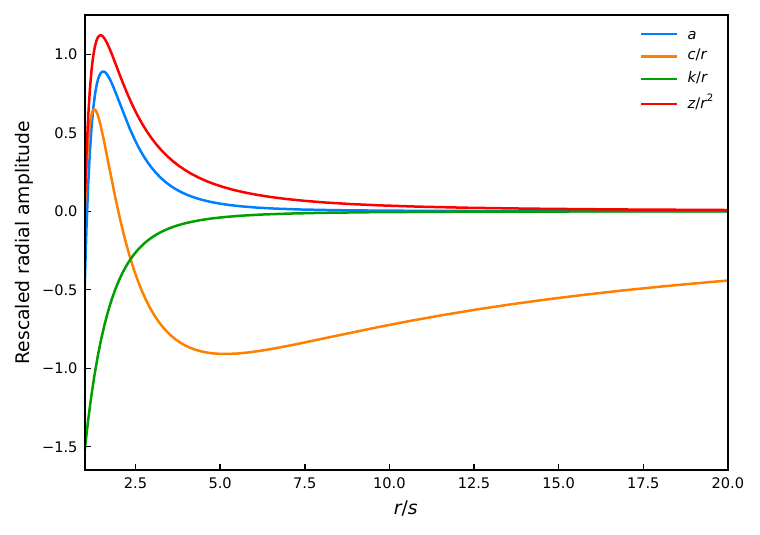}
\caption{Recomputed complementary static dipole mode at $\xi\simeq2.2931054$ for the reference parameters. The normalization is $\Phi_h=1$; $z_h\simeq0.04347$. The panels show the reconstructed amplitudes $a,z/r^2$ and $c/r,k/r$, respectively, with $k=K/U$. The horizontal axis is logarithmic.}
\label{fig:branchB_static_profile}
\end{figure}

The zero-frequency endpoint is determined directly from Eqs.~\eqref{eq:static_second_phi}--\eqref{eq:static_second_K}. Regularity imposes $c,K=\mathcal O(U)$ and leaves two horizon amplitudes. To specify the numerical boundary conditions explicitly, set $s=\lambda=1$, $L=\ell(\ell+1)$ and introduce
\begin{equation}
 f=\frac{3\Phi}{r},\qquad j=3r^2K.
\end{equation}
These variables should not be confused with the gauge-invariant amplitude $F=-\mu\Phi$. Their equations are
\begin{align}
 f'&=\frac{2f}{r}-\frac{3c}{rU},\\
 c'&=\frac{\xi(1-A)f}{r^2}+\frac{P_0z}{Er^4},\\
 z'&=\frac{(L-3\xi/r)j}{3Ur^2},\\
 j'&=\frac{5j}{r}-\frac{P_0Lf}{Er}.
\end{align}
At $r=1+\epsilon$, the leading regular conditions are
\begin{align}
 c&=\epsilon\left[\xi(1-A_h)f_h+P_0z_h/E_h\right]+\mathcal O(\epsilon^2),\\
 j&=-\epsilon P_0Lf_h/E_h+\mathcal O(\epsilon^2).
\end{align}
We fix $f_h=1$ in the computation and impose the leading decaying-physical-field conditions
\begin{equation}
 c=\frac23 f,\qquad j=\frac{P_0L}{5E_\infty}f
 \quad\text{at }r=r_{\max}.
\end{equation}
Terms omitted in these finite-radius conditions are tested by increasing $r_{\max}$. For the reference dipole, the collocation residual is below $10^{-8}$ and the outer-boundary comparison gives Table~\ref{tab:static_branchB_convergence}.
\begin{table}[tb]\centering
\caption{Complementary static dipole root at $\lambda=\bar P_0=1$, $\bar\gamma=-0.30$. The initial horizon offset is $10^{-5}s$.}
\label{tab:static_branchB_convergence}
\begin{tabular}{cc}\hline\hline
$r_{\max}/s$ & $\xi_0^{\rm static}$\\\hline
$1000$ & $2.293105477$\\
$3000$ & $2.293105376$\\\hline\hline
\end{tabular}\end{table}
The tracked root is therefore
\begin{equation}
 \xi_0^{\rm static}\simeq2.2931054.
 \label{eq:xi0_static_appendix}
\end{equation}
Reducing the horizon offset from $10^{-5}s$ to $10^{-7}s$ at $r_{\max}/s=3000$ changes the eigenvalue from $2.293105376464$ to $2.293105375989$. Thus the displayed static eigenvalue is limited mainly by the outer boundary in these tests. This precision describes the cutoff agreement, not a rigorous error bound. The radial profiles are shown in Fig.~\ref{fig:branchB_static_profile}. Rescaling to $\Phi_h=1$ gives $z_h\simeq0.04347$. The coefficients $a,c,k,z$ are nonzero, while $h=G=d=0$. We plot $a$, $z/r^2$, $c/r$, and $k/r$ to remove elementary coordinate factors. The displayed amplitudes are finite at the horizon and decay at infinity; their relative sizes are not energetic weights.

\subsection{Cutoff convergence}
\label{app:cutoff_convergence}
For a finite-frequency test we choose branch B at $\xi=2.4$, with the same background parameters. The mode is purely imaginary, $\omega_B=i\Gamma_B$. Table~\ref{tab:rmax_convergence} varies the outer cutoff with $\epsilon=10^{-9}$, matching radius $r_m/s=3.5$, and integration tolerance $3\times10^{-11}$.
\begin{table}[tb]\centering
\caption{Outer-boundary convergence of branch B at $\xi=2.4$.}
\label{tab:rmax_convergence}
\begin{tabular}{cc}\hline\hline
$r_{\max}/s$ & $\Gamma_Bs$\\\hline
100 & 0.037360056139\\
200 & 0.037360506355\\
400 & 0.037360506414\\\hline\hline
\end{tabular}\end{table}
The last two cutoffs differ by less than $6\times10^{-11}$. This rapid convergence is expected away from zero frequency because the selected outgoing solutions decay exponentially.

Table~\ref{tab:horizon_convergence} instead holds $r_{\max}/s=400$ and varies the horizon offset. The leading residue initialization leaves a visible truncation error at $\epsilon=10^{-5}$, which is strongly reduced at the smaller offsets. We use
\begin{equation}
 \Gamma_Bs\simeq0.03736051
 \label{eq:GammaB_extrapolated}
\end{equation}
for the displayed reference frequency, without interpreting the root solver's residual as its total uncertainty.
\begin{table}[tb]\centering
\caption{Horizon-offset convergence at $\xi=2.4$, with $r_{\max}/s=400$.}
\label{tab:horizon_convergence}
\begin{tabular}{cc}\hline\hline
$\epsilon$ & $\Gamma_Bs$\\\hline
$10^{-5}$ & 0.037350893880\\
$10^{-7}$ & 0.037360438739\\
$10^{-9}$ & 0.037360506414\\\hline\hline
\end{tabular}\end{table}

\subsection{Finite-frequency approach to the branch-B zero mode}
\label{app:zero_crossing}
We follow branch B toward its static crossing from larger positive $\xi$. The frequencies in Table~\ref{tab:branchB_crossing} decrease toward zero. Because the decay length diverges as $\Gamma_B\to0$, the outer boundary must be increased in this test. The zero-frequency equations provide a better determination of the crossing than an assumed linear fit to a few finite-frequency points.
\begin{table}[tb]\centering
\caption{Branch-B growth rate near the static crossing. The horizon offset is $10^{-9}s$; the two columns compare outer boundaries.}
\label{tab:branchB_crossing}
\begin{tabular}{ccc}\hline\hline
$\xi$ & $\Gamma_Bs$ ($r_{\max}/s=1500$) & $\Gamma_Bs$ ($3000$)\\\hline
2.320 & 0.01313604749 & 0.01313604749\\
2.310 & 0.00919645797 & 0.00919645797\\
2.300 & 0.00452752043 & 0.00452752044\\
2.296 & 0.00219547311 & 0.00219548940\\\hline\hline
\end{tabular}\end{table}
The small but resolved cutoff shift in the last row illustrates this limitation. The static result $\xi_0\simeq2.2931054$ is distinct from the exact axial crossing at $\xi_{10}=2$. Following the other imaginary branch back from the collision gives $\Gamma_As\simeq0.13215113$ at $\xi=\xi_0$, fixing the branch ordering used in the main text.

\subsection{Extraction of the mode-merger point}
\label{app:merger_extraction}
On the imaginary axis, the locally chart-normalized matching determinant can be chosen real. We solve the double-root conditions
\begin{equation}
 \mathcal E(i\Gamma,\xi)=0,\qquad
 \partial_\Gamma\mathcal E(i\Gamma,\xi)=0
\end{equation}
with a centered difference for the second equation. Two endpoint and tolerance settings give Table~\ref{tab:merger_checks}. The change is about $4.1\times10^{-9}$ in $\xi_{\rm m}$ and $2.6\times10^{-9}$ in $\Gamma_{\rm m}s$. At the refined point the normalized matching matrix has smallest singular value below $3\times10^{-13}$.
\begin{table}[tb]\centering
\caption{Direct double-root computation. Coarse: $(r_{\max}/s,\epsilon,r_m/s)=(200,10^{-7},3.5)$, tolerance $2\times10^{-9}$, derivative step $10^{-5}$. Refined: $(350,10^{-9},4)$, tolerance $3\times10^{-11}$, derivative step $3\times10^{-6}$.}
\label{tab:merger_checks}
\begin{tabular}{ccc}\hline\hline
Setting & $\xi_{\rm m}$ & $\Gamma_{\rm m}s$\\\hline
Coarse & 2.858860848 & 0.162616487\\
Refined & 2.858860844 & 0.162616490\\\hline\hline
\end{tabular}\end{table}
We independently solve for complex frequencies on the larger-coupling side. For $\delta=\xi-\xi_{\rm m}>0$, Table~\ref{tab:offaxis_merger} gives the positive-real-part member; its partner follows from reality of the equations.
\begin{table}[tb]\centering
\caption{Computed off-axis frequencies near the corrected collision.}
\label{tab:offaxis_merger}
\begin{tabular}{ccc}\hline\hline
$\delta$ & $\operatorname{Re}(\omega s)$ & $\operatorname{Im}(\omega s)$\\\hline
0.0001 & 0.00088525745 & 0.16262573365\\
0.0004 & 0.00177026488 & 0.16265346092\\
0.0100 & 0.00881086105 & 0.16353651976\\\hline\hline
\end{tabular}\end{table}
Extrapolating $[\operatorname{Re}(\omega s)]^2/\delta$ from the first two points to $\delta=0$ gives $A_{\rm m}\simeq0.08853$. The third point shows the size of the departure from the leading square-root approximation over a larger interval. These calculations establish a double frequency root with off-axis continuation; a direct eigenvector-coalescence test is a separate question.

\subsection{Independent checks and scope of the numerical result}
\label{app:collocation_check}
The static collocation calculation and finite-frequency subspace matching solve different formulations. Their agreement at the branch-B endpoint is a useful cross-check. The analytic axial profiles provide another check: direct substitution of $(\ell,n)=(1,0),(1,1),(2,0),(1,2)$ into the corrected static equation gives identically zero residual. Their positive eigenvalues, rather than the old negative values, are essential to this test.

To identify a complementary $\ell=2$ static solution, we continue the angular coefficient from $L=2$ to $L=6$ in 41 steps. This gives $\xi=2.106274351$ at $r_{\max}/s=1000$ and $2.106274172$ at $3000$, with collocation residual below $10^{-8}$. An initial guess unrelated to this continuation can converge to a higher root near $11.44$. Accordingly, neither convergence of a single initial guess nor the existence of one exact axial mode proves completeness of the static spectrum.

The reference collision is followed continuously to each dipole static endpoint, and the off-axis roots are computed as complex roots, not assigned from a fitted square-root curve. Varying $\bar\gamma$ and $\bar P_0$ requires replacing the electric background in every coefficient of the system; the robustness tables use such separate calculations. These tests support the particular growing branches reported here. They do not establish the full damped spectrum, exclude additional unstable modes, or remove the constrained kinetic pathology described in Sec.~\ref{sec:kineticcheck}.

\end{document}